\documentclass[conference]{IEEEtran}
\usepackage{algorithm}
\usepackage{algpseudocode}
\usepackage{tabularx}
\usepackage{makecell}
\usepackage{xcolor}
\usepackage{soul}
\usepackage{textcomp}
\usepackage{graphicx}
\usepackage{amsmath,amssymb,amsfonts}
\usepackage{cite}
\usepackage{caption}
\usepackage{listings}
\usepackage{ragged2e}  
\usepackage{subcaption}
\def\BibTeX{{\rm B\kern-.05em{\sc i\kern-.025em b}\kern-.08em
    T\kern-.1667em\lower.7ex\hbox{E}\kern-.125emX}}

\begin{document}
\title{\textit{StickyInvoc}: Rethinking Task Models for High-throughput Workflows in the LLM Era}

\author{
    \IEEEauthorblockN{Thanh Son Phung and Douglas Thain}
    \IEEEauthorblockA{Department of Computer Science and Engineering\\
    University of Notre Dame, Indiana, USA\\
    Email: \{tphung, dthain\}@nd.edu}
}

\maketitle
\begin{abstract}
The integration of Large Language Models (LLMs) into high-throughput workflows is creating a new class of workloads on HPC clusters that promises to accelerate advances in scientific discovery with unprecedented generative capabilities. However, the traditional task model, designed for task isolation and fault tolerance, imposes a prohibitive overhead in this new domain: each task must create its computational state from scratch and destroy it upon completion. For each LLM inference task, this "create-destroy" model forces the repeated and costly transfer of multi-gigabyte model parameters from a long-term, reliable storage (e.g., distributed file systems) to a compute node's local disk, its CPU memory, and finally its GPU memory. This overhead, compounded by the inherently high startup cost of LLM inference, the typical scale of thousands of tasks in high-throughput workflows, and the heterogeneous and preemptible nature of high-throughput resources, presents a significant performance barrier.

To overcome this barrier, this paper presents \textit{StickyInvoc}: a symbiotic relationship between two new task models for high-throughput workflows. Specifically, a "sticky" task creates a persistent and inheritable state on a compute node from a user-provided template, but doesn't execute any goodput computation by itself. Instead, this state is then seamlessly inherited by subsequent "invocation" tasks, which perform the actual computation without incurring the state creation overhead or destroying the state upon exit. \textit{StickyInvoc}  thus allows the decoupling of the creation and destruction of computational states, allowing the computational state of LLM models (e.g., in GPUs) to be created once per sticky task
and its cost amortized over many subsequent invocation tasks. Our evaluation shows that when rewritten in the \textit{StickyInvoc} paradigm, a claim verification workflow consisting of 150k inferences achieves a 3.6x speedup on a stable testbed with 20 GPUs (10.4k to 2.9k seconds), and completes in just 784 seconds by incrementally scaling out to 186 otherwise idle GPUs (32.8\% of all GPUs in our cluster).

\end{abstract}

\section{Introduction}
\label{sec:intro}
\subsection{Background}

Large Language Models (LLMs)\cite{achiam2023gpt, team2024gemini, anthropic_claude_3} have emerged as a transformative technology, demonstrating a remarkable capacity for discerning intricate patterns within vast datasets, and are becoming exceptionally powerful tools across diverse scientific domains. Consequently, a \textbf{new class of HPC workloads} is on the rise that integrates lightweight LLM inferencing (typically with \textit{billions} of parameters) into traditional high-throughput workflows to accelerate the pace of scientific discovery. This trend, exemplified by LLM-backed advancements in protein folding\cite{wohlwendminifold,shah2024energy, vieira2024scaling} and distributed AI-driven scientific computing frameworks\cite{ward2021colmena,fan2024workflowllm,gao2025strategic}, introduces a novel and challenging workload class to HPC clusters.

\begin{figure}[t]
    \includegraphics[width=\columnwidth]{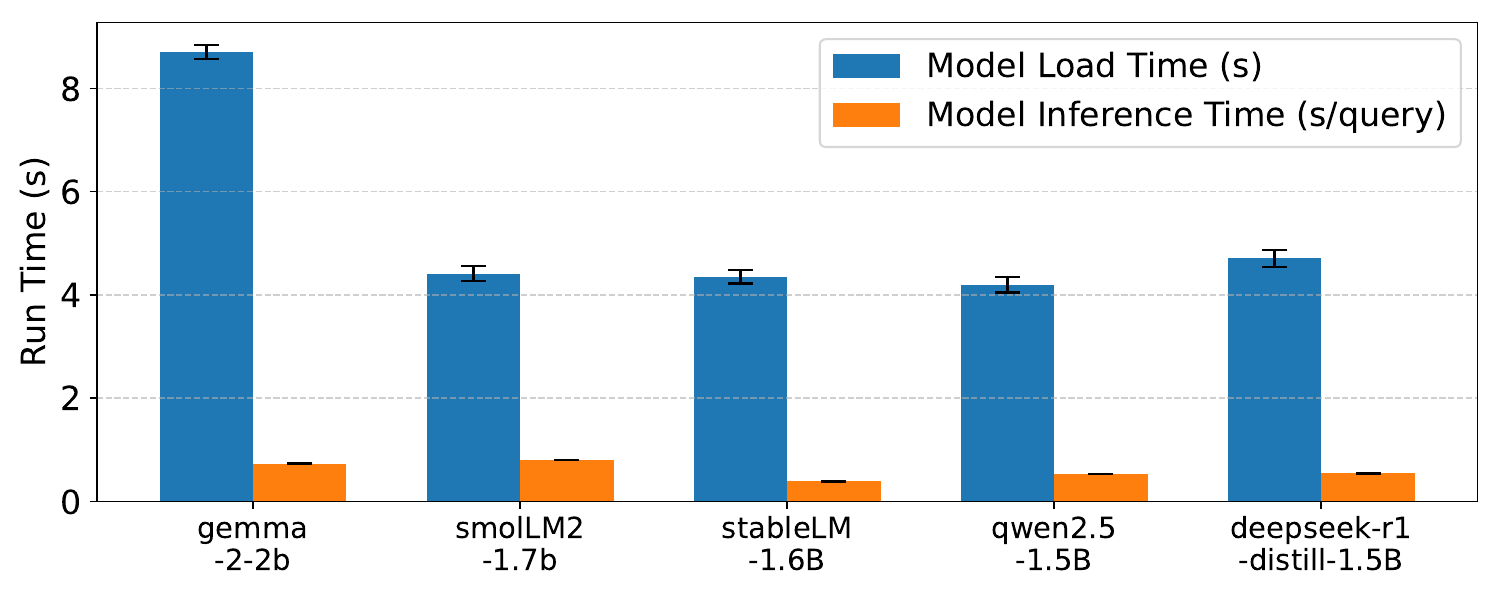}
    \caption{Model Loading and Inference Times across 5 Different LLMs on a Local Testbed (N=100). \textit{The local testbed includes an NVIDIA GeForce RTX 3060 GPU with 8 AMD Ryzen 7 5700G cores and a Micron M.2 2280 NVMe SSD as local storage. Each model loading and inference is repeated 100 times. In all cases, the model loading time dominates the overall inference computation, ranging from 84.6\% to 92.2\% of the total runtime.}}
    \label{plt:compare_5_models}
\end{figure}

The traditional task model in workflow systems \cite{deelman2015pegasus, radical, di2017nextflow, bui2011work}, a long-standing bedrock of high-throughput computing, defines a task as a \textbf{stateless, self-contained} execution unit. 

Specifically, each task is a discrete program that consumes inputs and produces outputs without relying on any pre-existing state on the execution node. This self-containment provides two key benefits: strong execution isolation and inherent fault tolerance. Isolation simplifies the concurrent execution of independent tasks and the development and deployment of complex workflows, while fault tolerance allows a failed task to be safely rescheduled and restarted on a different machine without data loss. Thus, this resilient model has been highly effective for a wide range of applications, from data-intensive bioinformatics pipelines \cite{wratten2021reproducible,ewels2020nf,van2020scalable} to large-scale simulations in high-energy physics \cite{sly2024reshaping, tovar2022dynamic, mehta2010pegasus}.

\subsection{Motivation}
\label{subsec:motivation}
\textbf{Overhead of LLM Inference Tasks.}  However, the stateless, self-contained nature of the traditional task model is fundamentally at odds with the computational behaviors of LLM inferences. Because a task must create its computational state from scratch, each LLM inference task must first load the model - consisting of gigabytes of parameters and associated software dependencies - before any useful computation can occur. This model loading process is a \textbf{costly, multi-stage data transfer}: the model is read from a long-term, reliable distributed filesystem over the network to a compute node's local storage, then loaded into CPU memory, and finally moved into GPU memory to be ready for inferences.
Figure \ref{plt:compare_5_models} shows the overhead of the same query inferred using 5 different lightweight LLMs \cite{gemma2b,smollm2, stablelm2,qwen2_5,deepseek-r1-distill} on an isolated testbed with all model parameters and software dependencies on a local NVMe SSD. Even with a localized setup involving no I/O over network, the model loading time in all cases consistently dominates the runtime to execute one query inference, varying from 84.6\% to 92.2\% of the total runtime. 

\begin{table}[t]
    \centering
    \begin{tabular}{|c|c|c|}
        \hline
         Device Name &  Release Year & Count \\
        \hline
               NVIDIA Quadro RTX 6000 & 2018 & 106\\
        \hline
        NVIDIA A10 & 2021 & 78 \\
        \hline
        NVIDIA TITAN X (Pascal) & 2016 & 69 \\
        \hline
        NVIDIA GeForce GTX 1080 Ti & 2017 & 63 \\
        \hline
        NVIDIA RTX 6000 Ada Generation & 2022 & 36 \\
        \hline
        NVIDIA GeForce GTX TITAN X & 2015 & 34 \\
        \hline
        NVIDIA A40 & 2020 & 26 \\
        \hline
        NVIDIA H100 80GB HBM3 & 2023 & 15 \\
        \hline
    \end{tabular}
    \caption{8 Major GPU Models in the Local Cluster. \textit{Our local HPC cluster contains 567 GPUs with 18 models in total, with 75\% of them in one of the 8 major models in the table, showing the heterogeneity of available GPUs and emphasizing the challenges of running LLM-integrated workflows on high-throughput resources.}}
    \label{tab:gpus}
\end{table}

Since the traditional stateless model mandates that all state be destroyed upon completion, this exorbitant startup cost is incurred for \textit{every} task. In a typical high-throughput workflow comprising thousands of tasks, this "create-destroy" cycle results in massive duplication of otherwise readily shareable model states between LLM inference tasks, and thus incurs a huge, yet avoidable as we shall demonstrate, performance degradation. 
Furthermore, the concurrent execution of many such tasks can create a "thundering herd" problem, overwhelming the distributed filesystem with simultaneous requests for the same directories containing model data and software dependencies. This contention not only strains the shared storage system but also adds a significant and variable time penalty to the already lengthy model loading process. 
The conventional approach to mitigate this overhead is to amortize the startup cost by batching multiple inference requests within a single task, thereby increasing the ratio of useful computation to setup time.

\textbf{The Chaotic Nature of High-throughput Resources.}
While batching inferences can be an effective strategy, its utility is constrained by the operational realities of HPC clusters. The standard practice of requesting fixed-size, exclusive resource allocations is often at odds with the longer execution times associated with large inference batches, as many HPC systems favor shorter jobs to improve resource access and job scheduling efficiency \cite{osc_batchprocessing,princeton_jobpriority,alcf_queuescheduling}. Compounding this issue is the current GPU market, where soaring demand and high prices \cite{Deloitte_2025,pilz2025trends,kachris2025survey} have led to heavily oversubscribed and contended job queues, resulting in significant wait times for users \cite{luo2024scheduling,liang2024resource,ding2023mirage}. Furthermore, the rigidity of these static allocations can lead to cluster-wide under-utilization due to resource fragmentation, leaving valuable GPU cycles idle \cite{jeon2018multi,xiao2020antman,jeon2019analysis}.

On the other hand, allocations on high-throughput resources that leverage otherwise idle components grants nearly immediate resource access with heavily discounted chargebacks \cite{nersc_policy,ncar_jobs,umd_queues} (typically 75\% to 100\% - free resource usage). However, this immediacy comes at the cost of stability and predictability: resources of any type can join and leave the idle pool at any moment, depending on the overall state of the cluster. Most clusters that provide access to idle resources do so in a \textbf{preemptible} manner\cite{utah_preemption,fnal_slurm,buffalo_jobs}: jobs are executed when idle resources become available and are evicted when higher-priority jobs arrive. 
Consequently, increasing the inference batch size per task, which linearly increases the task's runtime, also elevates the risk that the task will be preempted before completion, resulting in the loss of all accumulated computational work. This dynamic thus forces users to carefully tune the inference batch size per task to strike a delicate balance between amortizing startup costs and mitigating the risk of preemption.

This tuning process unfortunately adds another layer of complexity for non-technical users, and the optimal batch size, which depends on the startup cost and the preemption rate, is not straightforward to derive either, even for technical users. While the startup cost of an LLM inference task can be profiled on a local GPU, a typical cluster has many models of GPUs, reflecting the cluster's evolution over time as older hardware are gradually phased out and newer hardware are incrementally added in. For example, Table \ref{tab:gpus} demonstrates the heterogeneity of our local HPC cluster with 8 major GPU models spanning 8 years and accounting for 75\% of all GPUs in the cluster. This \textbf{heterogeneity} thus complicates the validity of a workflow's runtime profiling with a local GPU:  preemptible resources can come with \textit{any} GPU model, and an optimal batch size for one GPU model doesn't necessarily translate to optimality on other GPUs. Furthermore, the rate of resource preemption is completely \textit{dynamic and unpredictable over time} as it depends on, among other factors, the current load on the local cluster, the arbitrary resource demands from other jobs, and the specific allocation and scheduling policies of the cluster manager.

The combination of 4 challenges - the create-destroy overhead in the traditional task model, the inherently high startup cost of LLM inferences, the typical large scale of workflows, and the preemptible and heterogeneous nature of high-throughput resources - presents a significant performance barrier for the new class of high-throughput LLM-integrated workflows. This confluence of challenges leads to the following central research question of this paper: 
\textbf{How can we transform existing LLM-integrated workflows such that they can execute efficiently on high-throughput resources without repeatedly paying the startup cost upon preemption and/or incurring a high toll on users?}

\subsection{Limitation of State-of-the-art Approaches}
Existing methodologies for managing elastic and fault-tolerant computations are not well-suited to resolve the core tension between the high startup cost of LLM inferences and the volatile nature of high-throughput resources.

First, conventional autoscaling frameworks\cite{qiu2023aware,zou2024optscaler,augustyn2024tuning,catillo2023survey} are fundamentally mismatched with the preemptible resource model. Autoscaling systems operate on a \textbf{proactive} principle: the workflow itself initiates scale-up or scale-down events based on its internal workload, such as a rising queue of user requests. In the preemptible HPC setting, the workflow is purely \textbf{reactive}: it has no control over its allocated resources as they are allocated and preempted by the cluster manager based on external priorities. Therefore, a workflow cannot "request" more resources to meet demand, nor can it "release" them gracefully: it must simply adapt to the resources it is given, whenever they appear or disappear.

\begin{figure}[t]
    \includegraphics[width=\columnwidth]{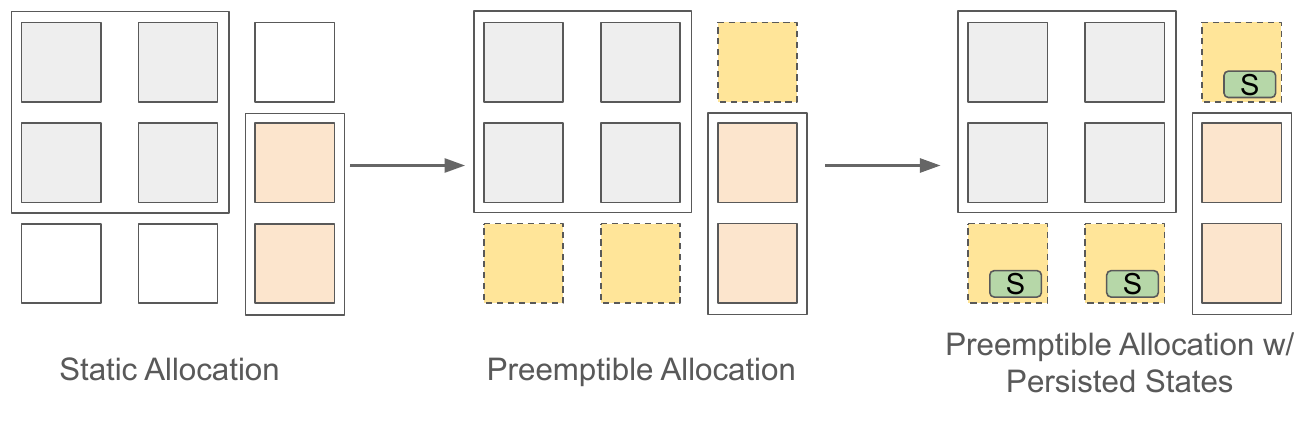}
    \caption{Persisted States in HPC Clusters. \textit{(Left) A cluster manager first prioritizes resource allocations to static jobs on standard batch queues. (Middle) Preemptible resources allow faster access to transiently available resources, but incurs a high startup penalty from LLM model initialization. (Right) Sticky tasks create persistent LLM states on preemptible resources such that invocation tasks can be readily matched and executed without incurring startup overheads.}}
    \label{fig:alloc-mode}
\end{figure}

Second, traditional fault-tolerance techniques like progress checkpointing \cite{goulart2023checkpointing, islam2012mcrengine, siachamis2024checkmate} offer only a partial and inadequate solution. While checkpointing the results of completed inferences allows a task to protect its progress, it does not address the primary problem: the prohibitive model loading cost. Upon preemption, a new task must still be instantiated on a different GPU, incurring the full startup penalty before it can resume work from the last checkpoint.

\subsection{Key Insights and Contributions}
Our approach is founded on a key insight: the primary performance bottleneck is not the computation but the tight coupling of inference execution with expensive model initialization. We propose to \textbf{decouple} these elements by 
introducing \textit{StickyInvoc}: a symbiotic relationship between two new task models for high-throughput workflows.

Specifically, we introduce a "sticky" task that creates a persistent and inheritable state on a compute node from a user-provided template, but doesn't execute any goodput computation by itself. This state is deliberately held on remote resources for subsequent "invocation" tasks to inherit, which perform the actual
computation without incurring the state creation overhead or
destroying the state upon exit.
When an invocation task is preempted from one node, it is simply requeued and rapidly rescheduled to another node that already holds the required  state. 
Furthermore, when new nodes join the resource pool, they can receive an existing state template directly from another node, cutting down the data transfer time and preventing a bottleneck at the distributed filesystem. 
This symbiotic relationship between sticky and invocation tasks effectively decouples the traditional create-destroy paradigm and makes the high cost of LLM initialization a one-time, amortizable expense per sticky task. It also alleviates the complex problem of searching for an optimal batch size as the startup cost is now shared across many invocation tasks. Figure \ref{fig:alloc-mode} visualizes on a high level how LLM states are persisted with sticky tasks on preemptible resources in HPC clusters.

Based on these ideas, this paper makes the following contributions:
\begin{enumerate}
    \item We analyzed the traditional task model in workflow systems and implemented an LLM-integrated claim verification workflow in the Parsl-TaskVine distributed data-intensive framework\cite{babuji2019parsl, sly2023taskvine, phung2023maximizing} with LLM inference tasks written in the traditional task model.
    \item We detailed a quick workflow transformation that decouples the model loading process from the actual inferences by transforming LLM inference tasks into sticky and invocation tasks. 
    \item We conducted a comprehensive evaluation demonstrating that this transformation significantly speeds up the end-to-end execution time of the workflow by 3.6x (from 10.4k seconds to 2.9k seconds), and allows it to scale up to 32.8\% of all 567 GPUs in the cluster and further reduces the execution time to 784 seconds.
\end{enumerate}
\subsection{Limitation of the Proposed Approach} 
The primary constraint of the proposed approach is that \textit{StickyInvoc} only applies to LLMs that are lightweight enough to fit within the resources of a single compute node (up to billions of parameters depending on the GPU setup per node). This is a direct consequence of the nature of high-throughput resources in HPC clusters, which are typically allocated and preempted on a per-node basis. Additionally, our system introduces its own management overhead for persisting states with sticky tasks, and its effectiveness is contingent on this overhead remaining substantially lower than the cost of repeated cold starts from a distributed filesystem.

\section{Implementation of an LLM-integrated Claim Verification Workflow}

\begin{figure}[t]
    \includegraphics[width=\columnwidth]{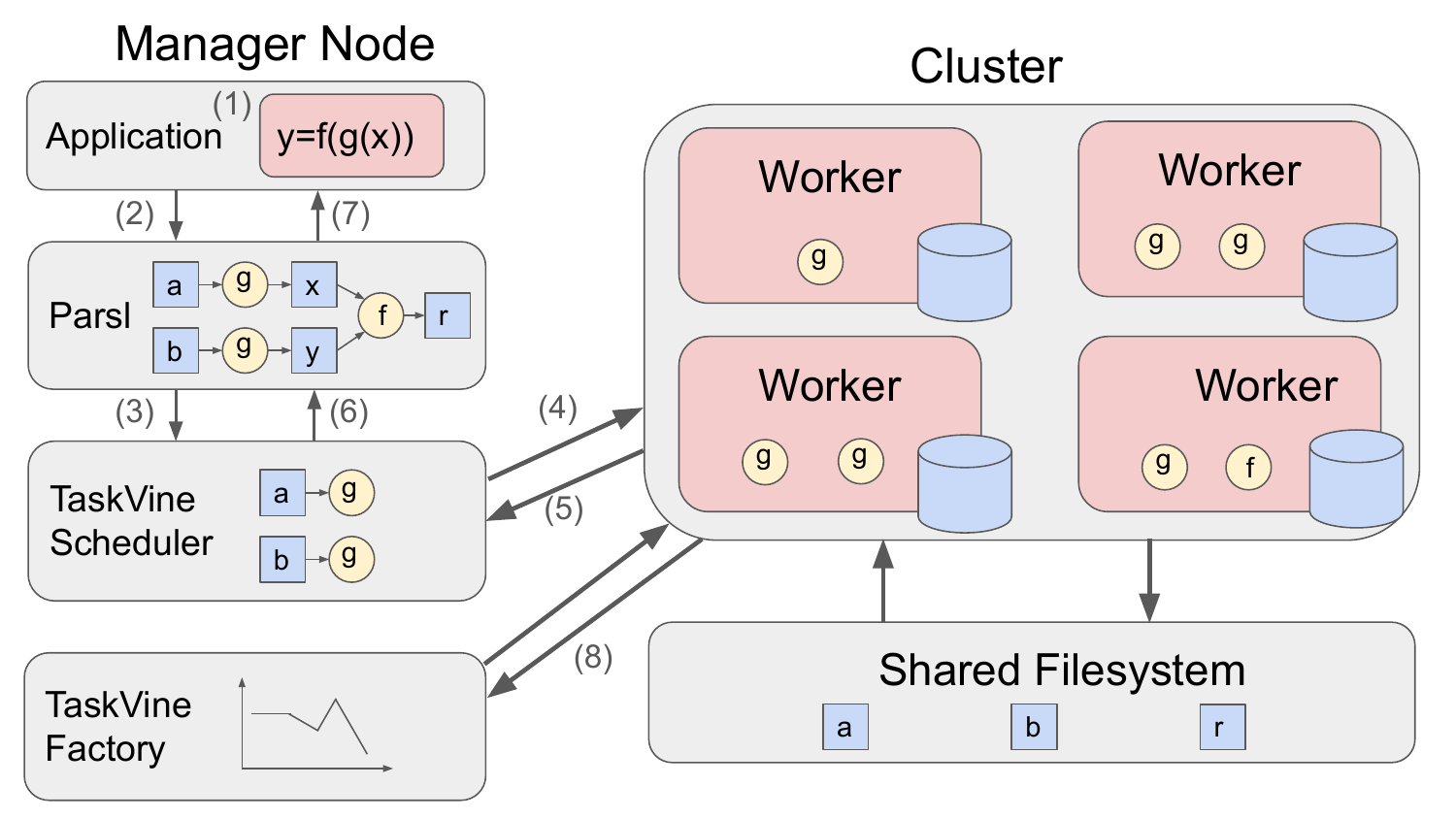}
    \caption{Overview of the Parsl-TaskVine Framework. \textit{The application defines computations via Python functions and passes them to Parsl. Parsl manages dependencies between functions and sends ready ones to the TaskVine scheduler. The scheduler manages resources on workers, schedules functions to available ones, and controls their execution and I/O patterns. The TaskVine factory monitors the connected resources and adjusts the quantity of workers accordingly.}}
    \label{fig:software-stack}
\end{figure}

\subsection{Overview of the Parsl-TaskVine Framework}
The underlying framework that powers the claim verification workflow is the integrated software stack of two dynamic workflow systems - Parsl\cite{babuji2019parsl} and TaskVine\cite{sly2023taskvine}. Parsl is a Python-native parallel library that allows users to express their computational needs via generic Python functions and automatically scales the computation on thousands of compute nodes, mainly focusing on flexibility, portability, and ease of use. 
TaskVine is a low-level data-intensive workflow execution engine that allows users to express low-level details about tasks and their inter-relationships. It then extracts values from the provided information to make intelligent scheduling and optimization decisions that accelerate large-scale data processing applications\cite{phung2024accelerating, phung2024adaptive}.

Figure \ref{fig:software-stack} shows how these two workflow systems work together in the big picture (the following descriptions correspond to numbers as denoted in the figure).
\textbf{(1)} On the manager node, a user expresses their  computational needs (e.g., LLM inferences) via generic Python functions. Once the workflow is run and these functions are invoked, \textbf{(2)} they are intercepted and passed to Parsl for inter-function dependency management and function-to-task translation. \textbf{(3)} Parsl sends ready tasks to the TaskVine scheduler, where they are examined for common execution and I/O patterns and scheduled for execution on workers accordingly. 
\textbf{(4)} The TaskVine scheduler manages resources in the system via TaskVine workers, where each worker is a small standalone pilot job that waits for instructions from the TaskVine scheduler and operates duly. \textbf{(5)} Once tasks are completed, workers communicate the results back to the scheduler, which \textbf{(6,7)} forwards them back to the application level. 
The TaskVine scheduler does not delegate the local resource management to individual workers: each task comes with a specific amount of resource allocation, and each worker is directed by the TaskVine scheduler on how to utilize any local resource type (CPU, memory, SSD, GPU). \textbf{(8)} The pool of resources is maintained by the TaskVine factory, a daemon-like process that monitors the current resource pool and adjusts it based on a given resource policy and the current load of the cluster.

\subsection{Implementation of the LLM-integrated Claim Verification Workflow}
Given this software stack, it is then straightforward for a user to implement a high-throughput LLM-integrated inference workflow.
A user first defines an arbitrary computation involving LLM inferences in a Python function. This function then flows through Parsl and the TaskVine scheduler to a TaskVine worker as a task to be executed. 
Each worker is allocated with a small number of GPUs such that a given task can run comfortably. Once the task completes, inference results are sent from the worker back to the application as described above. The scheduler has a queue of ready tasks, and its main job is to occupy connected workers with tasks at any given time. Therefore, the workflow will make progress as long as there are workers connected to the scheduler.

Additionally, this software stack provides a seamless integration with high-throughput resources. The scheduler on the manager node directs all workers on what to do and thus keeps a globally consistent view of the workflow. This means that workers can leave and join the pool freely as tracked by the TaskVine scheduler and adjusted by the TaskVine factory, and any preempted task is detected, retrieved, and re-inserted into the queue of ready tasks by the scheduler.

Figure \ref{fig:sample-code-context-agnostic} shows a simplified implementation of an inference function of the claim verification workflow (we describe the details of this workflow in Section \ref{subsec:settings}). To minimize the toll on non-technical users when scaling up a local application into a large-scale workflow, Parsl provides a decorator ("@python\_app") that wraps around a typical inference function. Remote execution is triggered when the user invokes the inference function as usual, allowing Parsl to intercept the inference invocation and prepare it for remote execution. The inference is then given to TaskVine for scheduling and execution, and the result is transparently sent back to the application. Alternatively, a user can instruct the local execution to wait for the result by calling the ".result()" method before continuing its execution flow.

Notice that this implementation uses the traditional "create-destroy" model for LLM inference tasks, and thus forces an inference function to reload its LLM state whenever it is preempted from a high-throughput resource as it couples the expensive model loading process with the actual inference execution into one executable unit (i.e., a task). The next section shows how we can decouple these two computational elements into sticky and invocation tasks to allow the efficient reuse of a one-time model startup cost per sticky task over multiple invocation tasks.

\section{Transforming the Workflow to Enable \textit{StickyInvoc} in LLM Inferences}

\subsection{Code Transformation to Enable \textit{StickyInvoc} in the LLM-integrated Claim Verification Workflow}
\label{subsec:transform}

We first show how the \textit{StickyInvoc} transformation looks like via a code sample, and then describe the implementation of \textit{StickyInvoc} with sticky and invocation tasks.
Since the startup cost of initializing an LLM is expensive, a workflow should define it as a state template, an input to the sticky task. When newly available resources arrive to the resource pool, the manager sends the sticky task to the new nodes to materialize the state templates into actual states and retained for subsequent reuses. The inferences then are defined as invocation tasks which produce the actual goodput. As multiple inferences arrive as invocation tasks to the task queue, the manager sends them to nodes that already have the state initialized for immediate inference execution. When resources are preempted by the cluster, these invocation tasks are seamlessly requeued by the manager for execution on other nodes that already host the needed state, eliminating the need to reinitialize the LLM from scratch per invocation task.

Figure \ref{fig:sample-code-context-aware} shows a 
code example of how the claim verification workflow can be quickly transformed to benefit from this technique. We first decouple, or split, the previous inference function into two new functions: "load\_model" and "infer\_model". Lines 3-7 define the "load\_model" function that creates an LLM state by loading its parameters from disk to GPU and returns this state via a dictionary to the StateManager process (this process manages the materialized states on remote nodes and is described further in Subsection \ref{subsec:implement}). This dictionary informs the StateManager of the relevant state to be exposed later to the actual inference invocation. Lines 9-14 define the actual computation via the "infer\_model" function that inherits the model directly from the initialized state held by the StateManager (instead of loading it from scratch), executes the inferences, and returns the results. Lines 18-24 connect the missing pieces of the example where the state template is defined via the \verb|parsl_spec| variable, and "infer\_model" brings this template reference along with its inputs to the scheduler for remote execution.

\subsection{Implementation Overview of \textit{StickyInvoc} in the Parsl-TaskVine Framework} 
\label{subsec:implement}

\begin{figure}[t]
\begin{lstlisting}[language=Python, basicstyle=\ttfamily\scriptsize, numbers=left, numberstyle=\tiny\color{gray}, numbersep=3pt, xleftmargin=8pt]
from parsl import python_app
# `infer` in the traditional task model
@python_app
def infer(model_path, claims):
    ...
    # this code loads the LLM
    model = AutoModel.from_pretrained(model_path)
    model.to('gpu')
    # inferences execute when the LLM is in the GPU
    verdicts = [model.generate(c) for c in claims]
    return verdicts
model_path = ...
claims = ...
verdicts = infer(model_path, claims).result()
\end{lstlisting}
\caption{Code Example of an LLM-integrated Claim Verification Workflow. \textit{An inference function is annotated with a Parsl-provided decorator, and remote execution is triggered by invoking the function as usual. Note that the traditional task model requires the coupling of the model loading process to the actual inference execution.}}
\label{fig:sample-code-context-agnostic}
\end{figure}

\begin{figure}[t]
\begin{lstlisting}[language=Python, basicstyle=\ttfamily\scriptsize, numbers=left, numberstyle=\tiny\color{gray}, numbersep=3pt, xleftmargin=8pt]
from parsl import python_app
# `load_model` is an input to the sticky task
def load_model(model_path):
    ...
    model = AutoModel.from_pretrained(model_path)
    model.to('gpu')
    return {'model': model}
# `infer_model`, when invoked, is an invocation task
@python_app
def infer_model(claims, parsl_spec):
    from parsl import load_variable_from_state_manager
    model = load_variable_from_state_manager('model')
    verdicts = [model.generate(c) for c in claims]
    return verdicts
model_path = ...
claims = ...
# inputs to the sticky task are specified here
parsl_spec = {'template': {'fn': load_model,
                          'args': [model_path],
                          'kwargs': {}}}
# an invocation task binds to the sticky task
# to later inherit relevant states via the 
# `parsl_spec` variable
verdicts = infer_model(claims, parsl_spec).result()
\end{lstlisting}

\caption{Code Example of a StickInvoc-enabled LLM-integrated Workflow. \textit{The previous inference function is now broken down into two new functions: "load\_model" that creates the model state in the GPU, and "infer\_model" that inherits and reuses the existing model state and runs inferences. "infer\_model" specifies its model template as an argument, and the remote execution is triggered as usual.}}
\label{fig:sample-code-context-aware}

\end{figure}

\begin{figure}[t]
    \includegraphics[width=\columnwidth]{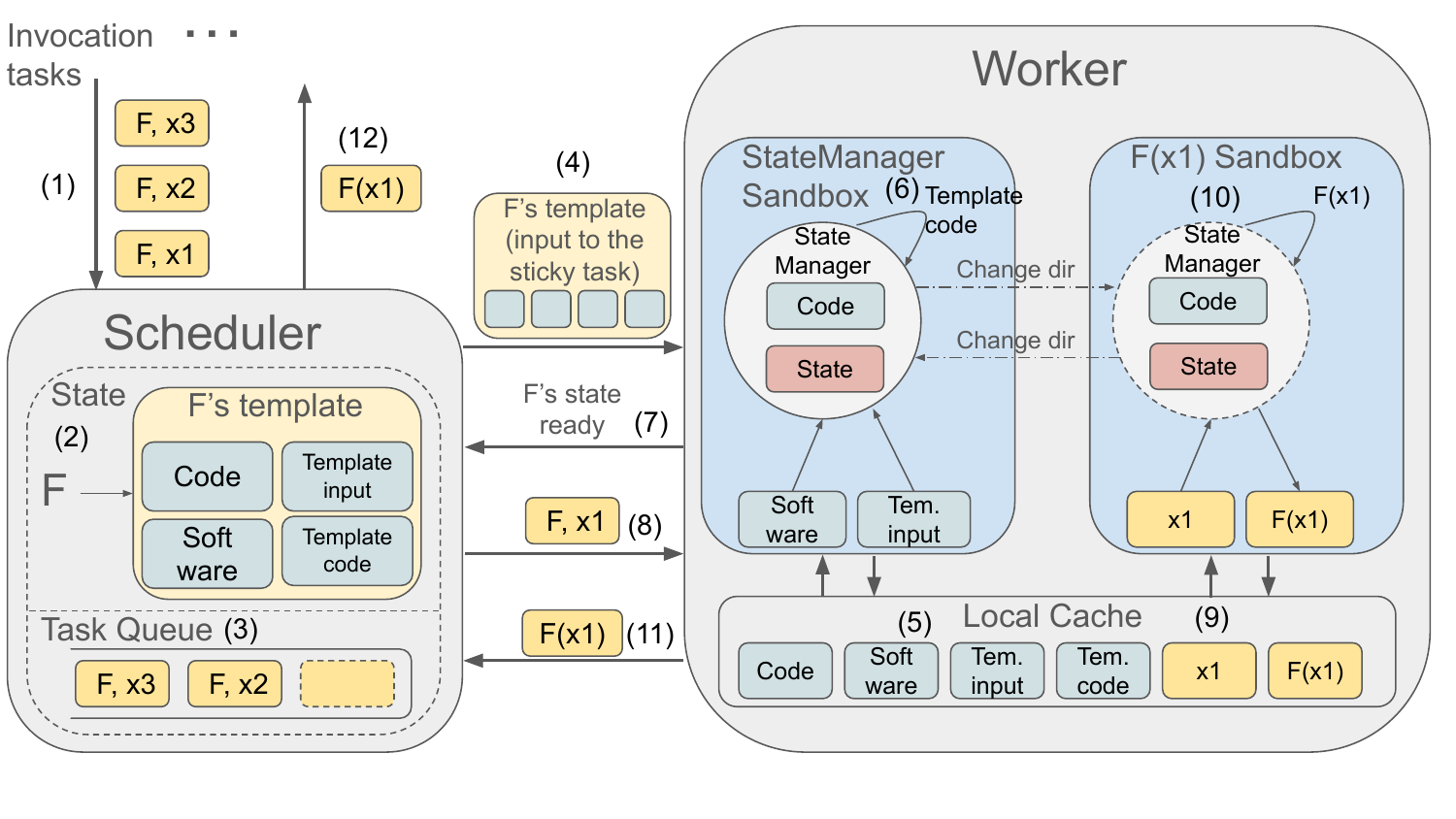}
    \caption{Implementation Overview of \textit{StickyInvoc}. \textit{The TaskVine scheduler analyzes F for its template upon the first invocation request, and sends it to the worker. The sticky task produces the StateManager process in the worker, which  registers F's code and creates F's state from the template and persists it locally. This state and registered code are then used to execute the current invocation tasks, and subsequent invocations inherit and reuse this existing state to speed up their executions.}}

    \label{fig:detailed-context-mgmt}
\end{figure}

Figure \ref{fig:detailed-context-mgmt} demonstrates how a computational state is persisted with a sticky task and reused by subsequent invocation tasks without repeatedly creating their own states in the Parsl-TaskVine stack (the following descriptions correspond to numbers as denoted in the figure). \textbf{(1)} An application (not shown) starts up and invokes a given function F repeatedly with different arguments (e.g., x1, x2, and x3 respectively). Parsl (not shown) sees that these function invocations don't have any dependency between them and converts them into ready tasks to be sent to the TaskVine scheduler. \textbf{(2)} The scheduler examines F and discovers its template to be shared between invocations of F, including F's code, software dependencies, template code, and template inputs,
\textbf{(3)} enqueues ready invocation tasks, and \textbf{(4)} sends the sticky task and this common template as its input to be materialized into a state and persisted on a given worker as part of the execution of the first invocation (e.g., F with x1). \textbf{(5)} The worker, upon receiving the template, stores all of its components in a local cache and \textbf{(6)} fork-execs the sticky task, resulting in a "StateManager" process. This process is responsible for materializing and persisting F's live state from its template and will cooperate with the worker to execute subsequent invocations of F. 
Upon the stage-in of F's template into its sandbox, the StateManager registers F's code, executes the template code, stores the resulting state internally in its process, and \textbf{(7)} lets the worker know it's ready for invocations of F. \textbf{(8)} The scheduler, upon receiving 
this ack from the worker, sends the first invocation task of F with x1. 
\textbf{(9)} The worker stores x1 in its cache, creates a sandbox for the invocation, and pings the StateManager. \textbf{(10)} The StateManager then changes its working directory to F(x1)'s sandbox and executes the invocation directly in its address space, which already contains F's state, before returning to its sandbox. \textbf{(11)} The result of the invocation is then returned to scheduler, which marks the completion of F(x1) and \textbf{(12)} forwards the result back to the application. Subsequent invocations of F (e.g., F-x2 and F-x3) then reuse F's already live state in the StateManager and follow the same path (F, x1) took.

It's important to note that the sticky task, which materializes into the StateManager process upon execution, is implemented using the traditional "create-destroy" task model. Thus, when a worker node or the StateManager process fails, the TaskVine scheduler automatically detects the failure and quickly schedules a new sticky task to a new node. When this happens, the state template, treated as a data dependency of a sticky task, is transferred from another node already hosting the template to the new node. This is to avoid the "thundering herd" problem on the distributed filesystem and implemented using the peer transfer feature in the TaskVine scheduler. Furthermore, unlike a traditional task which produces output files upon execution, the sticky task instead produces a StateManager process that cooperates with the worker process. Finally, the scheduler reclaims resources assigned to sticky tasks upon workflow termination or when it needs resources immediately and sticky tasks aren't being actively used.

\section{Evaluation}
\label{sec:eval}

This section begins with an in-depth description of the claim verification workflow along with the general experiment settings that apply to all evaluation efforts. Our evaluation then aims to answer the following research questions:
\begin{itemize}
    \item \textbf{RQ1 - Workflow Performance on the Stable Testbed}. How well does the workflow perform before and after the \textit{StickyInvoc} transformation on the stable testbed?
    \item \textbf{RQ2 - Runtime Analysis of LLM Inference Tasks.} How well does the \textit{StickyInvoc} transformation help reduce the model loading cost of individual LLM inference tasks?
    \item \textbf{RQ3 - Workflow Sensitivity to Varying Inference Batch Sizes}. How does the \textit{StickyInvoc} transformation help users pick the right inference batch size for the workflow?
    \item \textbf{RQ4 - Workflow Performance with Aggressive Resource Preemption}. How well does the \textit{StickyInvoc}-transformed workflow handle aggressive resource preemption from the cluster manager?
    \item \textbf{RQ5 - Workflow Performance with High-throughput Resources}. How well does the \textit{StickyInvoc}-transformed workflow scale when the capacity of transiently available resources in the cluster fluctuates?
\end{itemize}

\subsection{Experiment Settings}
\label{subsec:settings}
\textbf{Workflow.}
Claim verification is an active area of research given the lightning rise of online mis- and dis-information\cite{zhang2023towards, gunjal2024molecular}.
Our workflow, \textit{PromptVerify}, aims to find the \textit{optimal} prompt for a given LLM where it is used as a claim verifier to check the correctness of an arbitrary claim. 
Specifically, we use the training data from FEVER\cite{Thorne18Fever} as our dataset containing 145,449 claims, each of which is labeled with either \verb|SUPPORTED|, \verb|REFUTED|, or \verb|NOT ENOUGH INFO|. Each claim contains a statement about a given subject and a list of references to relevant Wikipedia pages.
Per the LLM, we use the recently released SmolLM2 model with 1.7 billion parameters\cite{allal2025smollm2}.
Our workflow takes the LLM and a prompt template, runs a full inference sweep across the dataset, and returns the claim verification accuracy. Note that the \textit{StickyInvoc} transformation is \textit{not} limited to a specific model or workflow and is applicable to all high-throughput LLM-integrated workflows subject to the limitation as described in Section \ref{sec:intro}.

\textbf{Local cluster.}
Our local cluster manages all nodes with HTCondor\cite{thain2005distributed} as the cluster manager. 
There are 567 GPUs in total in the cluster with 18 different GPU models (see Table \ref{tab:gpus} for 8 major  models). 
Our cluster provides access to data via the Panasas ActiveStor 16\cite{shaffer2017taming,panasas} distributed filesystem with 77 nodes and supports up to 84 Gbs/s read bandwidth and 94k read IOPS. 

\textbf{Parsl-TaskVine framework}. We configure parameters of our Parsl-TaskVine software stack as follows. We enable the peer transfer feature that allows workers to communicate and send arbitrary data between each other, which allows workers to send state templates in a peer-to-peer fashion and bootstrap newly arrived workers. Each task's resource allocation includes 2 cores, 10 GBs of memory, 20 GB of disk, and 1 GPU, providing a comfortable amount of resources for a smooth inference execution. Each TaskVine worker has 2 cores, 10 GBs of memory, 70 GBs of disk, and 1 GPU, thus providing the worker with just enough resources to run tasks in a 1-to-1 manner to preserve claimed preemptible resources and plenty of disk space for local caching.

Finally, almost all experiments start with the same resource pool configuration consisting of 20 GPUs, where half are NVIDIA A10 and the other half are NVIDIA TITAN X (Pascal). This approach allows us to not only establish consistency and stability to our measurements and results but also mimic the heterogeneity of the actual cluster (see Table \ref{tab:gpus}).
This constraint is removed at the end which allows the workflow to have access to up to
186 otherwise idle GPUs.
Storage-wise, the LLM takes up 3.7 GBs of disk and around 7.4 GBs of memory when fully loaded. The workflow's software dependencies are managed in a Conda\cite{conda} environment, containing 308 packages and totalling 10.5 GBs of disk.

\subsection{RQ1 - Workflow Performance on the Stable Testbed}
\label{subsec:rq1}
To quantify the impact of the \textit{StickyInvoc} transformation on the \textit{PromptVerify} workflow, we implement the traditional version of the workflow using the regular "create-destroy" task model for LLM inference tasks, and transform it into two other versions: \textit{StickyI/O} and \textit{StickyInvoc}. We detail the differences between these versions as below:

\begin{figure}[t]
    \centering
    \includegraphics[width=\columnwidth]{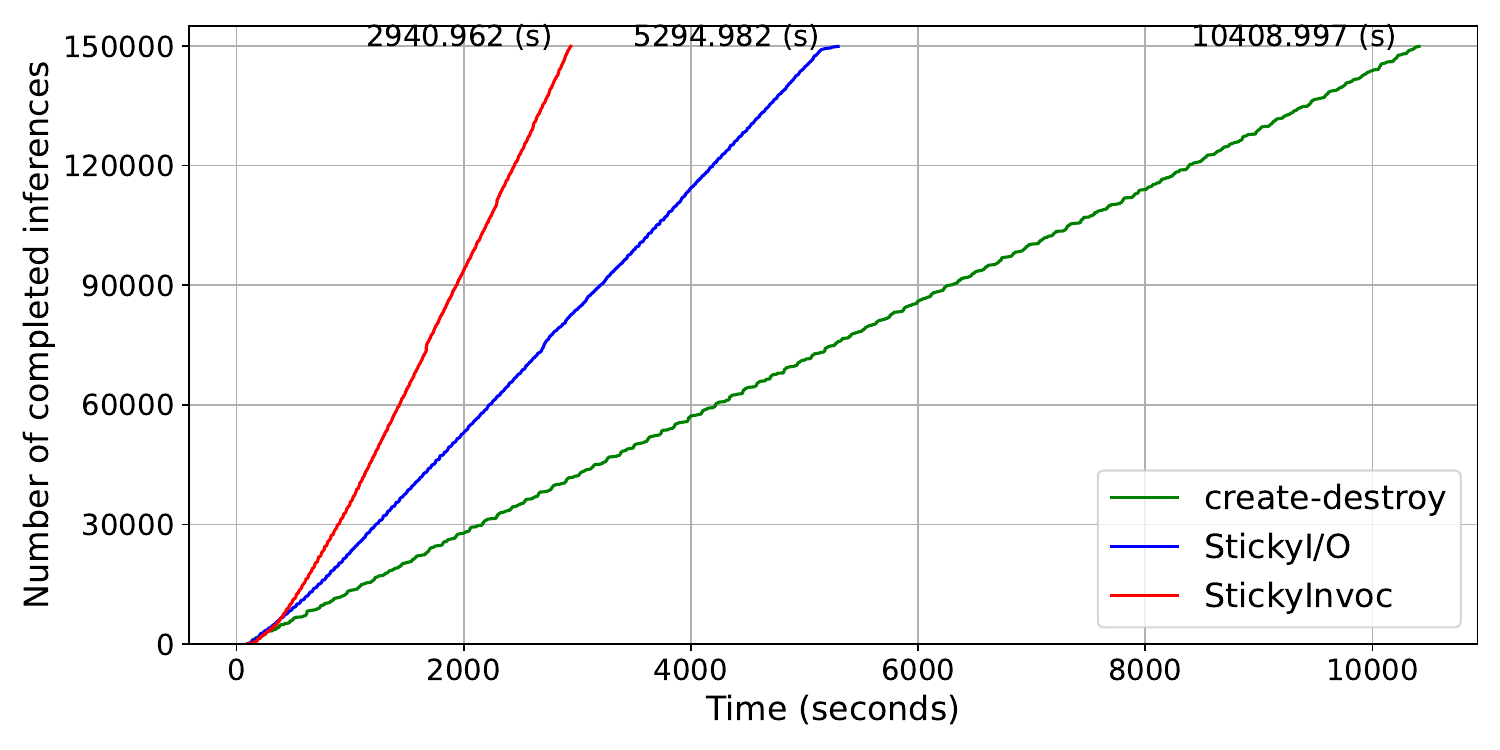}
    \caption{Execution Time of the \textit{PromptVerify} Workflow between Three Implementation Versions on Static Resources. \textit{The workflow is run with 3 different implementation versions: create-destroy (no state template is encoded in the workflow, forcing an LLM state initialization per task), StickyI/O (LLM state is persisted only on local disk of remote nodes, which includes GBs of model parameters and software dependencies), and \textit{StickyInvoc} (the LLM state further includes the model loading process in the local GPU for quick reuse). The end-to-end execution time of the workflow reduces drastically with "stickier" implementation versions.}}
    \label{fig:3_runtimes}
\end{figure}

\begin{itemize}
    \item \textbf{Create-destroy}. In the \textit{create-destroy} version, an LLM inference task makes all I/O calls for the model parameters and software dependencies to the distributed filesystem and creates a fresh state of the LLM model from scratch, involving moving GBs of data from the distributed filesystem to the local disk of the execution node, to its CPU memory, and finally to its GPU memory.
    \item \textbf{StickyI/O}. Instead of transferring GBs of input data per inference task, \textit{StickyI/O} caches these common input data on the local disk of  remote nodes so that subsequent tasks that run inferences on the same model can reuse the data available locally instead of making remote I/O calls to the distributed filesystem. Each task still has to create a fresh state of the LLM model in the GPU however.
    \item \textbf{StickyInvoc}. This version fully unlocks the benefits of the \textit{StickyInvoc} transformation and persists the LLM model state in the GPU with sticky tasks for efficient state reuse and fast state transfer between successive executions of invocation tasks. In other words, the LLM model state is created and loaded to the GPU once over a sticky task's lifetime.
\end{itemize}
All workflows are run on a stable testbed with statically allocated resources to isolate the results from other noises (e.g., fluctuations in capacity of high-throughput resources). Each task carries 100 inferences (i.e., inference batch size of 100), resulting in 1,500 tasks per workflow execution.

\begin{figure}[t]
    \centering
    \includegraphics[width=\columnwidth]{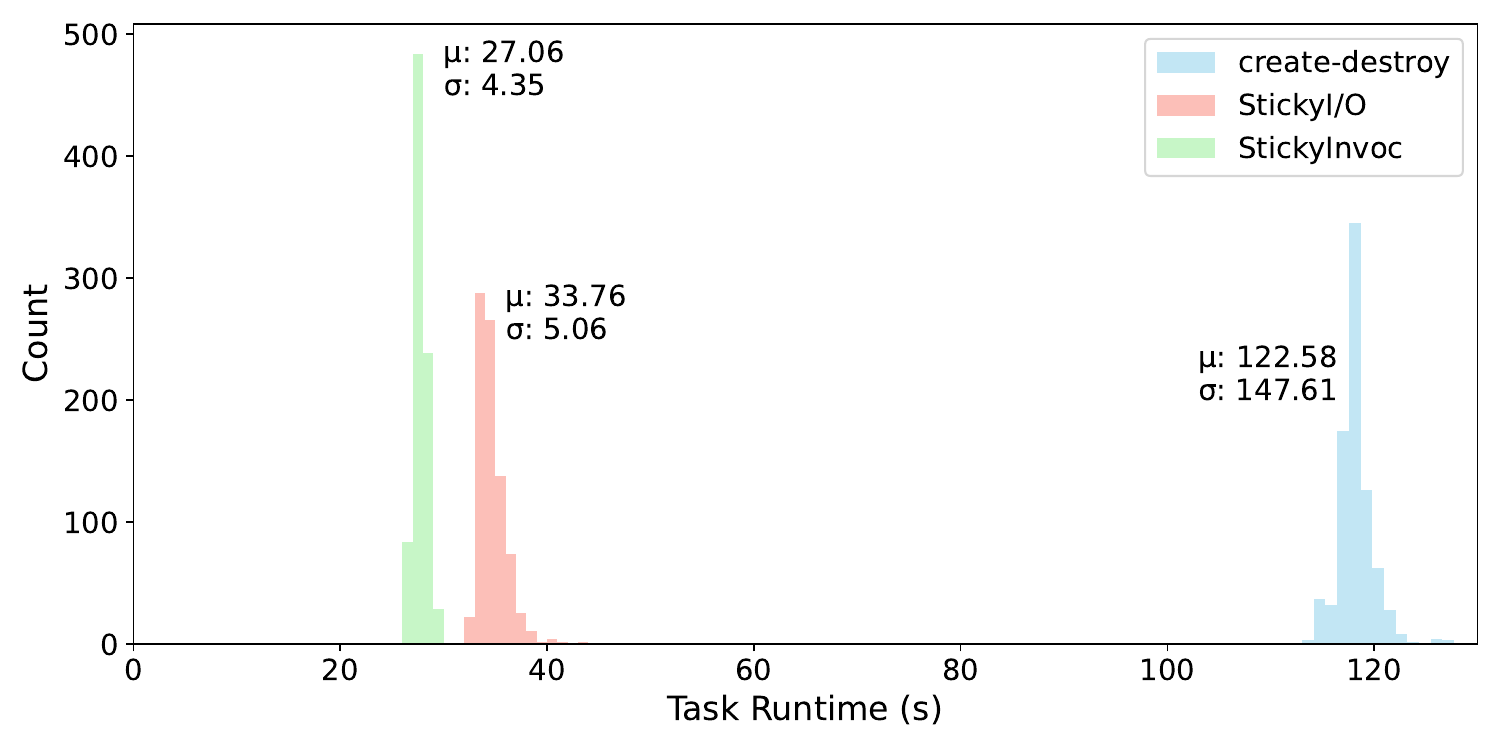}
    \caption{Histograms of Inference Task Runtimes for Three Implementation Versions. \textit{This figure shows the histograms of inference task runtimes of \textit{PromptVerify} with three implementation versions. As inference tasks are transformed and get stickier, the histogram of task runtimes shifts to the left (tasks on average run faster) and is thinner (the runtime deviations between tasks are smaller - tasks' runtimes are more stable.)}}
    \label{fig:histograms_runs}
\end{figure}

Figure \ref{fig:3_runtimes} shows the end-to-end execution time of 3 versions of the \textit{PromptVerify} workflow, each totaling 150,000 inferences. The result aligns with our expectation as the version with a "stickier" state has a significantly lower execution time. Specifically, the \textit{create-destroy} version runs the longest, over 10.4k seconds, as each task has to repeatedly make remote I/O requests to load GBs common input data from the distributed filesystem and construct a new GPU state of the LLM model upon initialization. \textit{StickyI/O} instead caches GBs of common input data on local disk of remote nodes and effectively converts most remote I/O requests into local ones. This thus helps bring the execution time down to 5.3k seconds, a significant speedup of the end-to-end execution time of 1.7x. \textit{StickyInvoc} cuts down the execution time even further to 2.9k seconds, a speedup in execution time of 3.6x and 1.9x compared to \textit{create-destroy} and \textit{StickyI/O}, respectively. This is because it eliminates both a bulk of unnecessary remote I/O requests for common input data and the need for an inference task to reconstruct the LLM state in a GPU upon startup as a task can now reuse an available state already initialized in a worker.

\begin{figure*}[t]
\begin{subfigure}[t]{0.325\textwidth}
\includegraphics[scale=0.35]{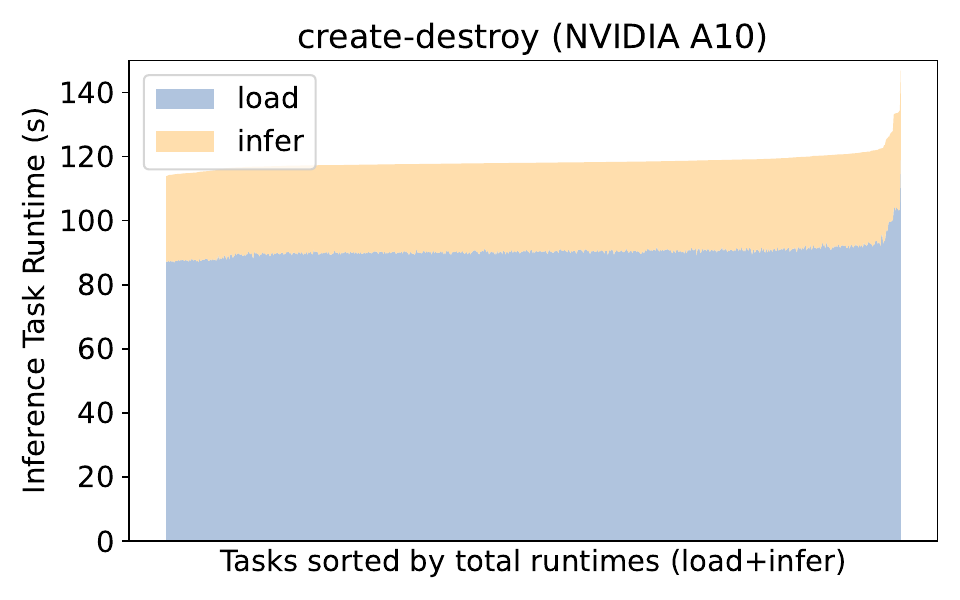}

\end{subfigure}
\begin{subfigure}[t]{0.325\textwidth} 

\includegraphics[scale=0.35]{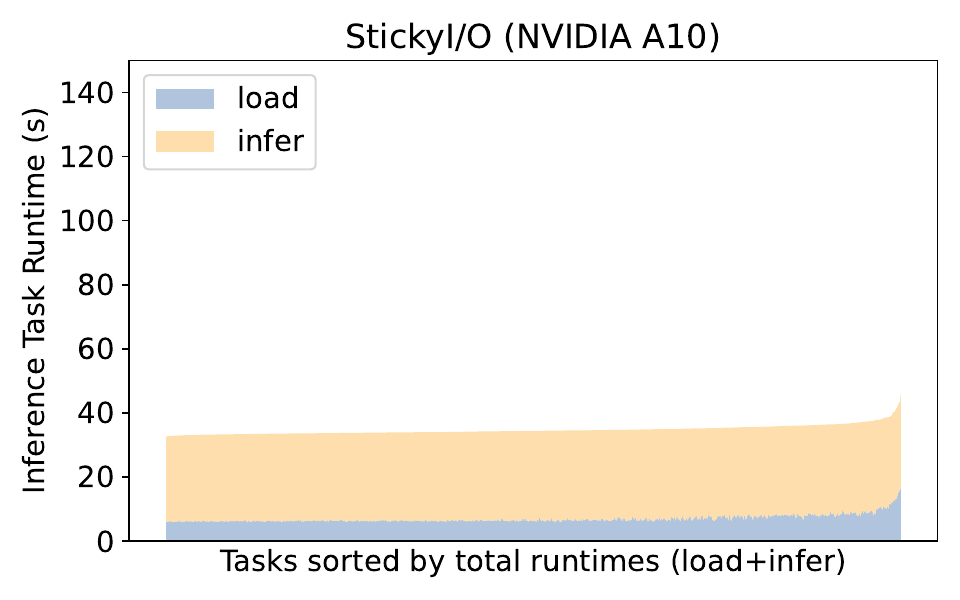}

\end{subfigure}
\begin{subfigure}[t]{0.325\textwidth}
\includegraphics[scale=0.35]{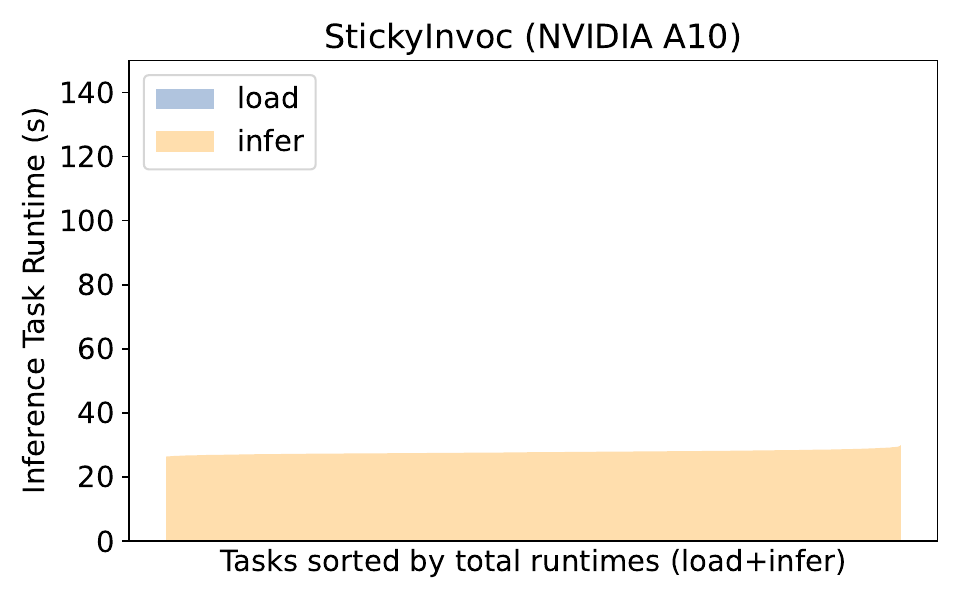}

\end{subfigure}

\vspace{0.4em}

\begin{subfigure}[t]{0.325\textwidth}
\includegraphics[scale=0.35]{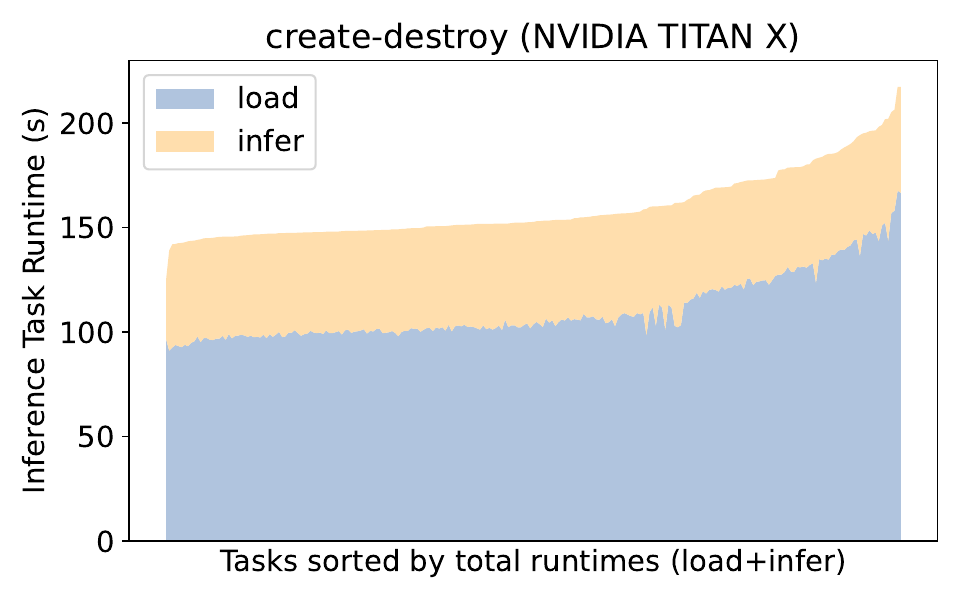}

\end{subfigure}
\begin{subfigure}[t]{0.325\textwidth}
\includegraphics[scale=0.35]{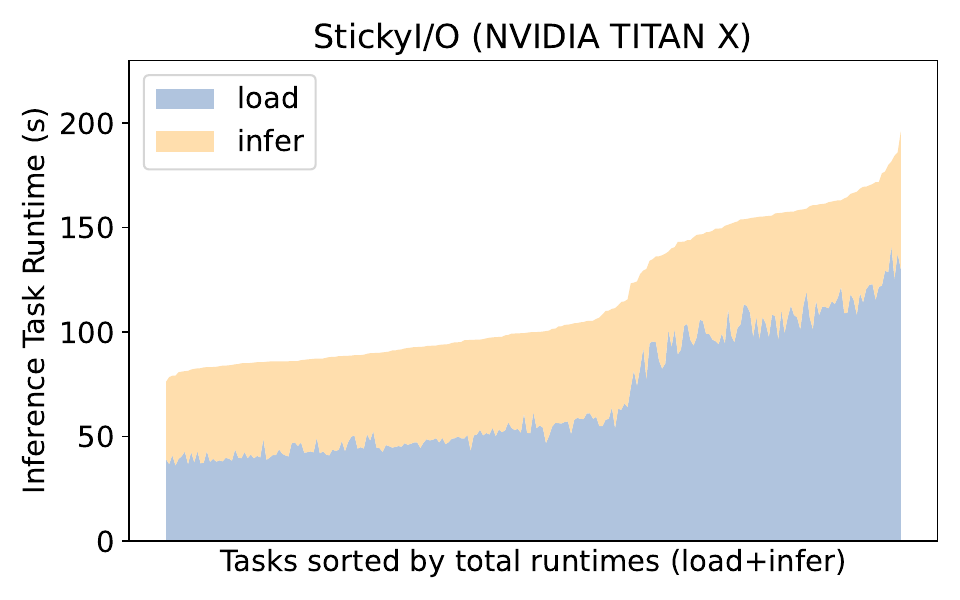}

\end{subfigure}
\begin{subfigure}[t]{0.325\textwidth}
\includegraphics[scale=0.35]{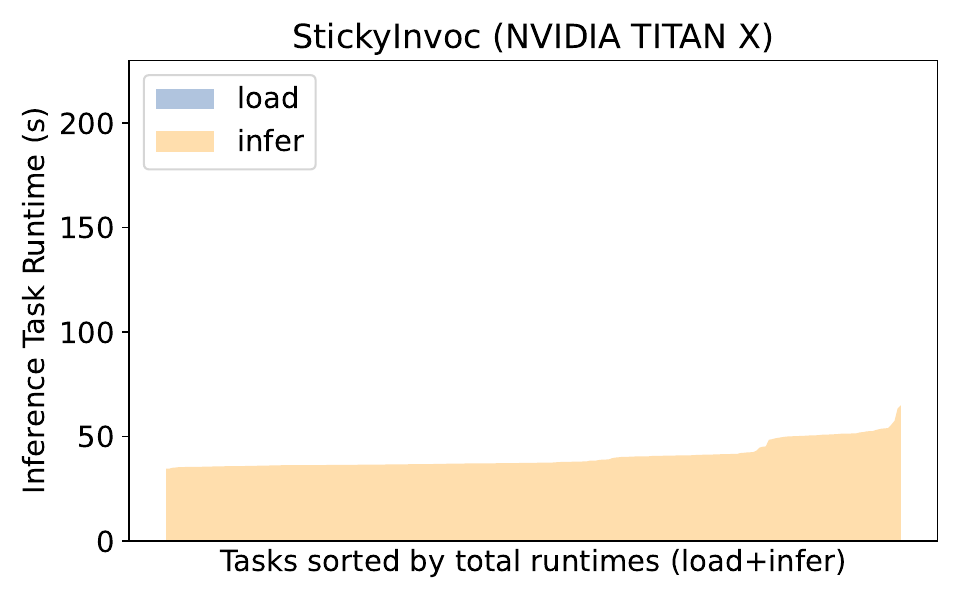}

\end{subfigure}

\caption{Breakdown of Inference Task Runtimes between 3 Implementation Versions (the top row shows results from tasks run on NVIDIA A10 GPUs, and the bottom row shows those run on NVIDIA TITAN X GPUs). \textit{(Left) The \textit{create-destroy} model forces each inference task to load its own model parameters from the distributed filesystem, creating a massive amount of remote I/O and state duplication. (Middle) \textit{StickyI/O} caches model parameters and software dependencies on remote nodes, which significantly reduces the model loading time. (Right) \textit{StickyInvoc} further persists the created model state in GPUs between subsequent invocation tasks, virtually removing the model loading cost.}}

\label{plt:runtime-breakdown}
\vspace{-.15in}
\end{figure*}

We then conclude that StickInvoc transformation of the \textit{PromptVerify} workflow allows a considerably more efficient execution with a huge reduction in the end-to-end execution time on the stable testbed.

\subsection{RQ2 - Runtime Analysis of LLM Inference Tasks}
\label{subsec:runtime-analysis}
We now measure one level deeper and analyze the runtimes of LLM inference tasks between three different implementation versions as described in  Subsection \ref{subsec:rq1}. Figure \ref{fig:histograms_runs} shows the histograms of task runtimes for all three versions (this figure only shows results from inference tasks run on NVIDIA A10 GPUs for runtime consistency, as two GPU models have different inference speeds). A huge  runtime reduction is observed for tasks in the \textit{StickyI/O} version compared to the \textit{create-destroy} version. 
This is because the model loading time is cut down significantly by caching and reusing LLM parameters and software dependencies on remote node's disk rather than every read operation going to the distributed filesystem. \textit{StickyInvoc} further persists the GPU state in memory and moves the histogram even further to the left. 
This is because each LLM inference task no longer needs to be loaded from disk to GPU memory per inference task, as the state is persisted using sticky tasks and reused by state inheritance on invocation tasks.
Note that the histograms are also thinner: the runtime standard deviation of \textit{StickyInvoc} is the lowest as the tasks' runtimes become more consistent and predictable due to the removal of the variable model loading process, especially from a distributed filesystem.

Figure \ref{plt:runtime-breakdown} shows the breakdown of the model loading time and inference time for inference tasks in 3 implementation versions.
In both GPU models, we can see that, due to the isolation and fault-tolerance guarantee from the \textit{create-destroy} task model, each task must load its model parameters from scratch, which involve GBs of data being transferred via I/O calls over the network. This results in a massive duplication of LLM state creation, with the majority of time in each task spent on loading the model itself. The middle figures show that when these parameters are instead cached on compute nodes with the \textit{StickyI/O} transformation, the runtime of inference tasks is reduced significantly as most I/Os happen locally instead of over the network. Note that the difference in the model loading time of \textit{StickyI/O} between nodes equipped with NVIDIA A10s and those with NVIDIA TITAN Xs is due to the quality of their hardware (a 6-year difference, see Table \ref{tab:gpus}). Finally, the right plots with the \textit{StickyInvoc} transformation show that when the LLM state is persisted in the GPUs, the model loading time is virtually eliminated for all inference tasks in both GPU models. This thus shows the effectiveness of the \textit{StickyInvoc} transformation in supporting high-throughput LLM-integrated workflows, as the majority of tasks' computations are now goodputs.

\subsection{RQ3 - Workflow Sensitivity to Varying Inference Batch Sizes}
Subsections \ref{subsec:rq1} and \ref{subsec:runtime-analysis} demonstrate how the \textit{StickyInvoc} transformation enables an efficient execution of the \textit{PromptVerify} workflow and addresses the first part of the central research question as posed in Subsection \ref{subsec:motivation}. This subsection then addresses the second part of the question and shows how this technique makes it easy for users to pick an inference batch size without worrying about optimal and/or sub-optimal workflow execution.

\begin{figure}[t]
    \centering
    \includegraphics[width=\columnwidth]{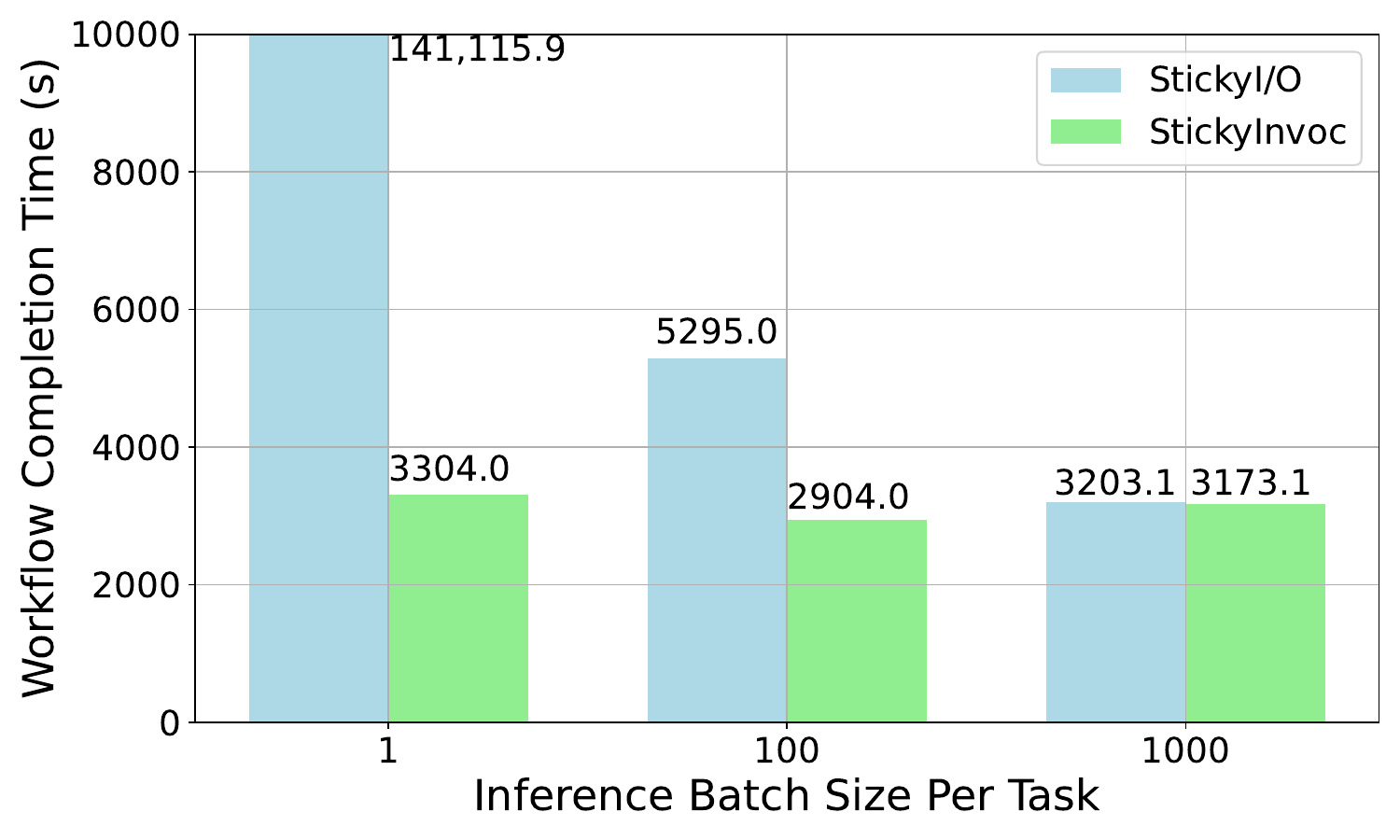}
    \caption{Effect of Inference Batch Size to the Workflow's Execution Time. \textit{StickyI/O and StickyInvoc are run with 3 different batch sizes: 1, 100, and 1000. StickyI/O introduces a large variance of execution time across batch sizes due to the expensive LLM model initialization cost per task, even with cached common input data. On the other hand, StickyInvoc stabilizes this range of execution time as each model is initialized once per GPU/sticky task and reused across multiple invocation tasks instead of once per regular task, alleviating the effect of a wrong choice of inference batch size.}}
    \label{fig:batch-size}
    \justifying
    \noindent
\end{figure}

Figure \ref{fig:batch-size} shows the execution time of \textit{StickyI/O} and \textit{StickyInvoc} with 3 batch sizes: 1, 100, and 1000 (we skip \textit{create-destroy} as it is clearly suboptimal as demonstrated in Subsection \ref{subsec:rq1}.) With a wrong inference batch size of 1, \textit{StickyI/O} takes a disastrous hit in its performance and needs 141.1k seconds to complete end-to-end. This is because each inference now needs to load the model from scratch, and the overhead of the model initialization completely dominates the total execution time. Even at the batch size of 100, \textit{StickyI/O} still takes 5.3k seconds and is still far from its best execution time with a batch size of 1000 at 3.2k seconds.

In stark contrast, \textit{StickyInvoc} delivers a remarkably stable execution time across all tested batch sizes.
The workflow runs the worst with an inference batch size of 1 at 3.3k seconds, and the best with batch size of 100 at 2.9k seconds. The range of execution time is thus limited to approximately 400 seconds, or 13.6\% of the best execution time, over the range of possible batch sizes of 1000 (from 1 to 1000). Such a stable range of execution time comes from the state persistence and reuse between tasks, and the performance difference is only the cumulative overhead of the state inheritance that happens once per invocation task. Thus, the \textit{StickyInvoc} transformation ensures that the workflow will not run disastrously with a wrong batch size and its execution is always optimal or near-optimal with any batch size, removing the worry of batch size tuning from users.

\subsection{RQ4 - Workflow Performance with Aggressive Resource Preemption}

High-throughput resources fluctuate frequently, increasing the capacity as more jobs exit and release their previously claimed resources, and decreasing the capacity as new jobs are allocated and scheduled by the cluster manager. This subsection focuses on the latter and quantifies how well \textit{StickyInvoc} executes the workflow efficiently with aggressive resource preemption.

\begin{figure}[t]
    \centering
    \includegraphics[width=\columnwidth]{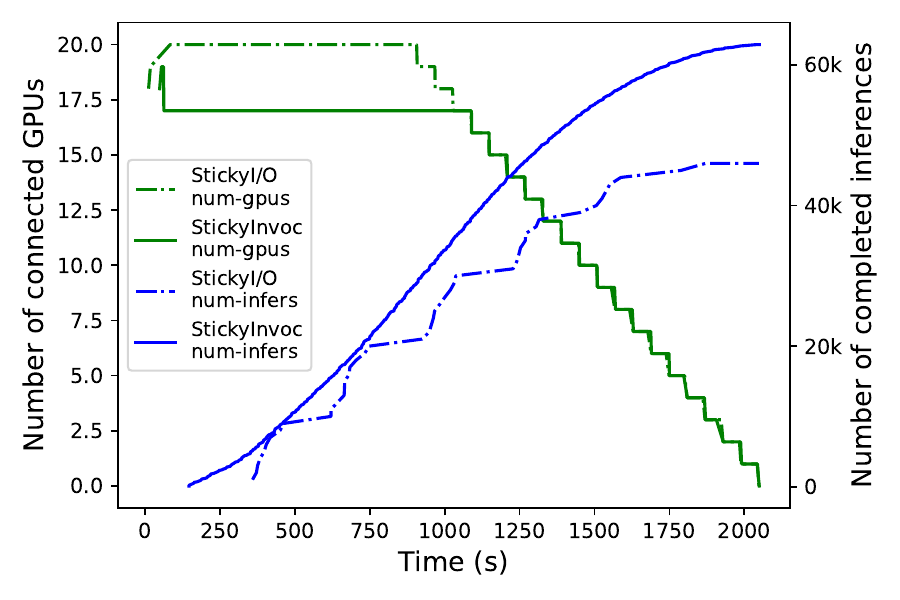}
    \caption{Number of Completed Inferences with Aggressive Resource Preemption. \textit{This figure shows the number of completed inferences over time between StickyI/O and StickyInvoc with aggressive resource preemption from the cluster (1 GPU preemption per minute). Despite an early drop of 3 GPUs, StickyInvoc still completes 16.9k inferences more than StickyI/O, and consistently has a higher inference completion rate at any given time.}}
    \label{fig:worker-preemption}
\end{figure}

Figure \ref{fig:worker-preemption} shows the scenario where resources are preempted aggressively from the workflow by the cluster manager with the preemption rate of 1 GPU per minute from the 900-second mark until the resource pool of the workflow is depleted (we preempt all NVIDIA A10s before NVIDIA Titan X Pascals). In this figure, we compare the number of completed inferences between \textit{StickyI/O} with its best inference batch size of 1000 and \textit{StickyInvoc} with that of 100 (see Figure \ref{fig:batch-size}). For \textit{StickyI/O}, we can see the "rugged" rate of inference completion that gradually flattens out from the resource depletion and ends with 46k completed inferences. This is due to the larger batch size of 1000 and the expensive LLM model initialization cost per task that blocks goodput until the LLM model is fully loaded in a GPU, creating the effect of low goodput when the LLM models are being loaded on remote workers, and high goodput when many inferences are executed.

\begin{figure*}[t]
\begin{subfigure}[t]{0.325\textwidth}
\includegraphics[scale=0.23]{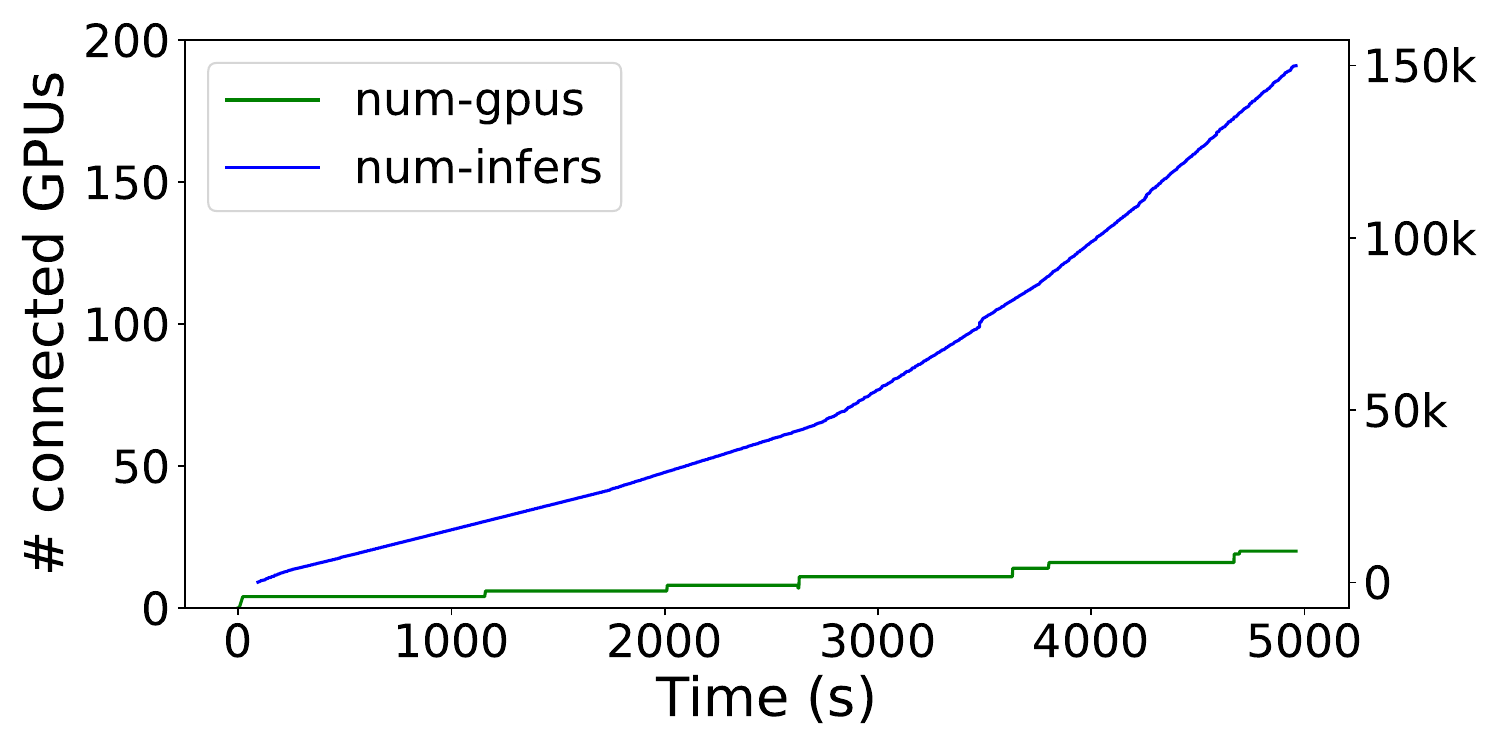} 
  \captionsetup{justification=centering}
  \caption{Low resource availability}
  \label{plt:low}
 \end{subfigure}
 \begin{subfigure}[t]{0.325\textwidth} 

 \includegraphics[scale=0.23]{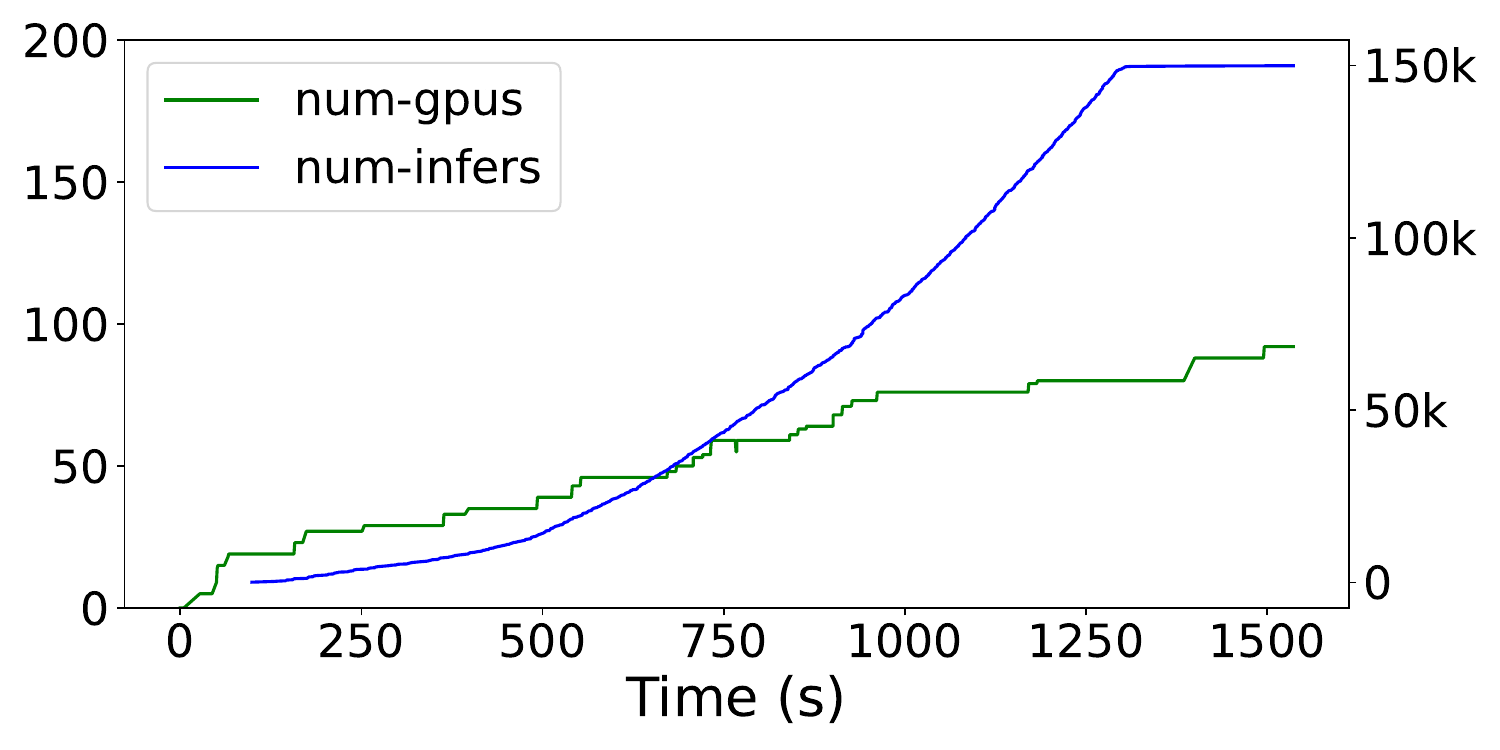} 
 \captionsetup{justification=centering}
  \caption{Medium resource availability}
  \label{plt:medium}
 \end{subfigure}
 \begin{subfigure}[t]{0.325\textwidth}
 \includegraphics[scale=0.23]{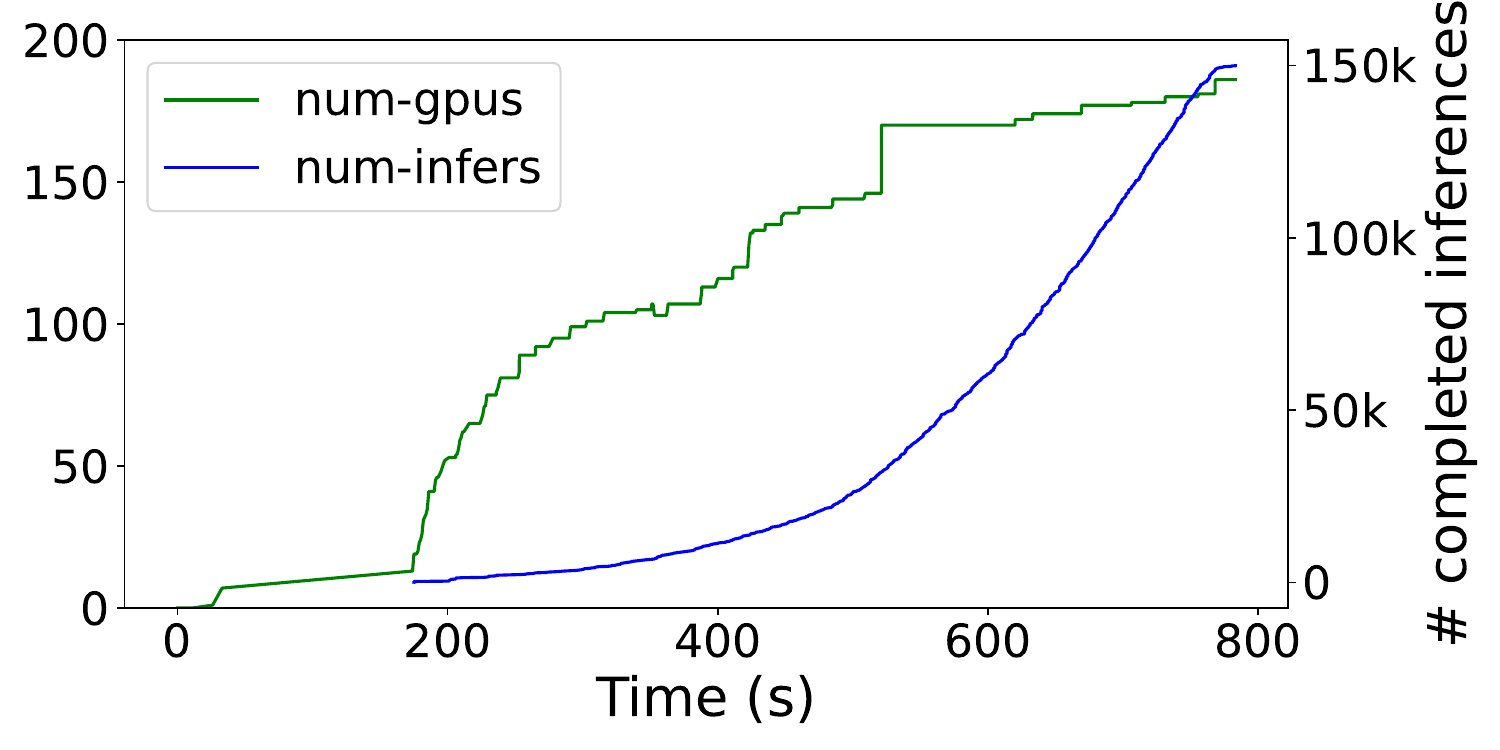}
 \captionsetup{justification=centering}
  \caption{High resource availability}
  \label{plt:high}
 \end{subfigure}

\caption{Workflow Resilience Against Dynamic Availability of High-throughput Resources. \textit{Note that plots share both y axes (number of connected GPUs and number of completed inferences over time) but have their own scales of the x axis (time in seconds). Workflow's inference progress seamlessly adapts to the availability of high-throughput resources (represented via the number of  connected GPUs) in all cases.}}

\label{plt:unres}
\vspace{-.15in}
\end{figure*}

Per \textit{StickyInvoc}, despite having an early drop of 3 GPUs due to external preemption from the cluster manager, it still completes 16.9k inferences more than \textit{StickyI/O} (ends with 62.9k completed inferences), and consistently has a higher inference rate than \textit{StickyI/O} (the number of inferences of \textit{StickyI/O} never exceeds that of \textit{StickyInvoc} at any point in time). Notice the smooth inference completion rate as it shows how tasks from preempted GPUs are seamlessly requeued and rerun with an already GPU-initialized LLM state, removing the rugged-like effect on the workflow's inference rate and allowing the slope to smoothly flat out at the end compared to that of \textit{StickyI/O}. Thus, the \textit{StickyInvoc} transformation helps the workflow smoothly make more progress even when resources are preempted aggressively.

\subsection{RQ5 - Workflow Performance with High-throughput Resources}
Finally, we focus on how well the StickyInvoc-transformed workflow scales with dynamic high-throughput resources in the local cluster. Figure \ref{plt:unres} shows the number of connected GPUs and the number of completed inferences over time for \textit{StickyInvoc} when the cluster has low (Subfigure \ref{plt:low}), medium (Subfigure \ref{plt:medium}) and high (Subfigure \ref{plt:high}) availability of high-throughput resources. In Subfigure \ref{plt:low}, the workflow only starts out with 4 GPUs and gradually goes to 20 GPUs, finishing in 4959 seconds. Note that even with a limited amount of GPUs, the workflow still makes consistent progress as the completed inference rate scales linearly with the amount of connected GPUs. Subfigure \ref{plt:medium} shows another instance of the workflow eventually claiming 92 GPUs, shortening the execution time to 1536 seconds. On the other hand, when the cluster has many jobs exiting and releasing their claimed GPUs (Subfigure \ref{plt:high}), the workflow quickly grabs up to 186 transiently available GPUs (32.8\% of all GPUs in the cluster) and finishes the execution in only 783 seconds. This thus shows that \textit{StickyInvoc}, a combination of sticky and invocation tasks, allows the \textit{PromptVerify} workflow to swiftly react and scale to the ever-changing amount of high-throughput resources in the cluster over time.

\section{Related Works}
\label{sec:related}

\subsection{Spot Instances} The use of underutilized compute capacity is a well-established practice in both commercial cloud computing and High-Performance Computing (HPC), though the implementation and guarantees vary. Cloud providers offer discounted Spot or Preemptible Instances\cite{GoogleCloudSpotVMs, AWSEC2Spot, AzureSpotVMs} that can be reclaimed at any time, similar to how preemptible resources are used in HPC. While several studies\cite{mao2025skyserve, miao2024spotserve} have explored running stateful applications like LLM inference on these cloud resources, they rely on a critical feature: a preemption warning\cite{AWSSpotTermination,GoogleCloudSpot,AzureSpotVMsDoc}. This notification period, typically 30 to 120 seconds, allows applications to checkpoint state or transfer work before termination. On the contrary, preemptible resources in many HPC environments offer no such warning, and preemption is instantaneous, rendering traditional state-saving mechanisms ineffective. Our work addresses this specific challenge by introducing \textit{StickyInvoc} designed to handle the abrupt and unpredictable nature of preemption in HPC clusters by retaining the common computational state between inferences in all connected GPUs.

\subsection{LLM Inference Optimization}
Many works optimize the inference process of huge LLMs by outputting several tokens in one forward pass based on the speculative decoding scheme\cite{leviathan2023fast,spector2023accelerating, chen2024cascade, svirschevski2024specexec}. This scheme assumes that an LLM generating tokens sequentially takes too much time and resources, especially with easy-to-predict tokens. To speed this up, a smaller LLM is used to predict the next K tokens in advance, and the original LLM can make one forward pass that accepts tokens it agrees with and rejects others instead of making K forward passes.
Other works focus on KV cache and memory management on both a single GPU and a pool of GPUs. Kwon et. al.\cite{kwon2023efficient} introduce a virtual paging mechanism that divides the dynamically-sized KV cache into blocks to remove GPU memory fragmentation, while Lin et. al.
\cite{lin2024infinite} distribute the KV cache and the attention computation to many GPUs.
Cloud deployment of inference serving is also an active area of research. Fu et. al.\cite{fu2024serverlessllm} use local storage of individual instances to cache and distribute model checkpoints among each other. Our work extends the usage of local storage to memory and GPUs to hold and distribute the computational state of lightweight LLMs. 

\subsection{Workflow Systems}
Workflow systems evolve from the traditional resource managers and allow applications to express complex relationships between tasks via a directed acyclic graph (DAG) instead of a bag of tasks\cite{deelman2015pegasus, turilli2019middleware, zheng2017deploying, phung2021not}. These systems typically focus on applications' reliability, performance, and portability via novel architectural designs and runtime optimizations, but require users to describe the computational needs in detail via complicated and non-uniform abstractions. More modern workflow systems\cite{babuji2019parsl, rocklin2015dask, moritz2018ray} tackle this usability problem by providing Pythonic abstractions that enable users to wrap their computational needs neatly into Python functions and translating these functions into tasks deployable to remote nodes. Our Parsl-TaskVine integration follows this movement and allows users to easily describe their computations in Python without losing performance, reliability, or portability. 
The Parsl-TaskVine stack extends this movement one step further with the support of computational state sharing between tasks on contrary to the traditional view of complete inter-task independence.

\section{Conclusion}

The traditional "create-destroy" task model is ill-suited for high-throughput LLM-integrated workflows, creating a performance barrier from repeated model initialization. This paper introduced \textit{StickyInvoc}, a new paradigm that decouples state creation from computation. By transforming a large-scale claim verification workflow, \textit{StickyInvoc} achieved a 3.6x speedup, showed resilience to aggressive preemption, and scaled to 186 opportunistic GPUs to finish in just 784 seconds. Our work shows that rethinking the task models enables LLM-integrated workflows to run efficiently and scalably on  HPC systems without imposing a heavy complexity toll on users.

\bibliographystyle{IEEEtran}
\bibliography{cclpapers, main}

\begin{thebibliography}{10}
\providecommand{\url}[1]{#1}
\csname url@samestyle\endcsname
\providecommand{\newblock}{\relax}
\providecommand{\bibinfo}[2]{#2}
\providecommand{\BIBentrySTDinterwordspacing}{\spaceskip=0pt\relax}
\providecommand{\BIBentryALTinterwordstretchfactor}{4}
\providecommand{\BIBentryALTinterwordspacing}{\spaceskip=\fontdimen2\font plus
\BIBentryALTinterwordstretchfactor\fontdimen3\font minus \fontdimen4\font\relax}
\providecommand{\BIBforeignlanguage}[2]{{%
\expandafter\ifx\csname l@#1\endcsname\relax
\typeout{** WARNING: IEEEtran.bst: No hyphenation pattern has been}%
\typeout{** loaded for the language `#1'. Using the pattern for}%
\typeout{** the default language instead.}%
\else
\language=\csname l@#1\endcsname
\fi
#2}}
\providecommand{\BIBdecl}{\relax}
\BIBdecl

\bibitem{achiam2023gpt}
J.~Achiam, S.~Adler, S.~Agarwal, L.~Ahmad, I.~Akkaya, F.~L. Aleman, D.~Almeida, J.~Altenschmidt, S.~Altman, S.~Anadkat \emph{et~al.}, ``Gpt-4 technical report,'' \emph{arXiv preprint arXiv:2303.08774}, 2023.

\bibitem{team2024gemini}
G.~Team, P.~Georgiev, V.~I. Lei, R.~Burnell, L.~Bai, A.~Gulati, G.~Tanzer, D.~Vincent, Z.~Pan, S.~Wang \emph{et~al.}, ``Gemini 1.5: Unlocking multimodal understanding across millions of tokens of context,'' \emph{arXiv preprint arXiv:2403.05530}, 2024.

\bibitem{anthropic_claude_3}
Anthropic, ``{The Claude 3 Model Family: Opus, Sonnet, Haiku},'' 2024, available at https://www-cdn.anthropic.com/de8ba9b01c9ab7cbabf5c33b80b7bbc618857627/Model\_\\Card\_Claude\_3.pdf.

\bibitem{wohlwendminifold}
J.~Wohlwend, M.~Reveiz, M.~McPartlon, A.~Feldmann, W.~Jin, and R.~Barzilay, ``Minifold: Simple, fast, and accurate protein structure prediction,'' \emph{Transactions on Machine Learning Research}, 2025.

\bibitem{shah2024energy}
A.~Shah and S.~Jayaratnam, ``Energy efficient protein language models: Leveraging small language models with lora for controllable protein generation,'' \emph{arXiv preprint arXiv:2411.05966}, 2024.

\bibitem{vieira2024scaling}
L.~C. Vieira, M.~L. Handojo, and C.~O. Wilke, ``Scaling down for efficiency: Medium-sized transformer models for protein sequence transfer learning,'' \emph{bioRxiv}, pp. 2024--11, 2024.

\bibitem{ward2021colmena}
L.~Ward, G.~Sivaraman, J.~G. Pauloski, Y.~Babuji, R.~Chard, N.~Dandu, P.~C. Redfern, R.~S. Assary, K.~Chard, L.~A. Curtiss \emph{et~al.}, ``Colmena: Scalable machine-learning-based steering of ensemble simulations for high performance computing,'' in \emph{2021 IEEE/ACM Workshop on Machine Learning in High Performance Computing Environments (MLHPC)}.\hskip 1em plus 0.5em minus 0.4em\relax IEEE, 2021, pp. 9--20.

\bibitem{fan2024workflowllm}
S.~Fan, X.~Cong, Y.~Fu, Z.~Zhang, S.~Zhang, Y.~Liu, Y.~Wu, Y.~Lin, Z.~Liu, and M.~Sun, ``Workflowllm: Enhancing workflow orchestration capability of large language models,'' \emph{arXiv preprint arXiv:2411.05451}, 2024.

\bibitem{gao2025strategic}
X.~Gao, Q.~Pei, Z.~Tang, Y.~Li, H.~Lin, J.~Wu, L.~Wu, and C.~He, ``A strategic coordination framework of small llms matches large llms in data synthesis,'' \emph{arXiv preprint arXiv:2504.12322}, 2025.

\bibitem{deelman2015pegasus}
E.~Deelman, K.~Vahi, G.~Juve, M.~Rynge, S.~Callaghan, P.~J. Maechling, R.~Mayani, W.~Chen, R.~F. Da~Silva, M.~Livny \emph{et~al.}, ``Pegasus, a workflow management system for science automation,'' \emph{Future Generation Computer Systems}, vol.~46, pp. 17--35, 2015.

\bibitem{radical}
M.~Turilli, V.~Balasubramanian, A.~Merzky, I.~Paraskevakos, and S.~Jha, ``Middleware building blocks for workflow systems,'' \emph{Computing in Science \& Engineering}, vol.~21, no.~4, pp. 62--75, 2019.

\bibitem{di2017nextflow}
P.~Di~Tommaso, M.~Chatzou, E.~W. Floden, P.~P. Barja, E.~Palumbo, and C.~Notredame, ``Nextflow enables reproducible computational workflows,'' \emph{Nature biotechnology}, vol.~35, no.~4, pp. 316--319, 2017.

\bibitem{bui2011work}
P.~Bui, D.~Rajan, B.~Abdul-Wahid, J.~Izaguirre, and D.~Thain, ``Work queue+ python: A framework for scalable scientific ensemble applications,'' in \emph{Workshop on python for high performance and scientific computing at sc11}, 2011.

\bibitem{wratten2021reproducible}
L.~Wratten, A.~Wilm, and J.~G{\"o}ke, ``Reproducible, scalable, and shareable analysis pipelines with bioinformatics workflow managers,'' \emph{Nature methods}, vol.~18, no.~10, pp. 1161--1168, 2021.

\bibitem{ewels2020nf}
P.~A. Ewels, A.~Peltzer, S.~Fillinger, H.~Patel, J.~Alneberg, A.~Wilm, M.~U. Garcia, P.~Di~Tommaso, and S.~Nahnsen, ``The nf-core framework for community-curated bioinformatics pipelines,'' \emph{Nature biotechnology}, vol.~38, no.~3, pp. 276--278, 2020.

\bibitem{van2020scalable}
B.~Van~de Sande, C.~Flerin, K.~Davie, M.~De~Waegeneer, G.~Hulselmans, S.~Aibar, R.~Seurinck, W.~Saelens, R.~Cannoodt, Q.~Rouchon \emph{et~al.}, ``A scalable scenic workflow for single-cell gene regulatory network analysis,'' \emph{Nature protocols}, vol.~15, no.~7, pp. 2247--2276, 2020.

\bibitem{sly2024reshaping}
B.~Sly-Delgado, B.~Tovar, J.~Zhou, and D.~Thain, ``Reshaping high energy physics applications for near-interactive execution using taskvine,'' in \emph{SC24: International Conference for High Performance Computing, Networking, Storage and Analysis}.\hskip 1em plus 0.5em minus 0.4em\relax IEEE, 2024, pp. 1--13.

\bibitem{tovar2022dynamic}
B.~Tovar, B.~Lyons, K.~Mohrman, B.~Sly-Delgado, K.~Lannon, and D.~Thain, ``Dynamic task shaping for high throughput data analysis applications in high energy physics,'' in \emph{2022 IEEE International Parallel and Distributed Processing Symposium (IPDPS)}.\hskip 1em plus 0.5em minus 0.4em\relax IEEE, 2022, pp. 346--356.

\bibitem{mehta2010pegasus}
G.~Mehta, E.~Deelman, K.~Vahi, and F.~Silva, ``Pegasus workflow management system: helping applications from earth and space,'' in \emph{AGU Fall Meeting Abstracts}, vol. 2010, 2010, pp. IN41B--1362.

\bibitem{gemma2b}
\BIBentryALTinterwordspacing
Google, ``Gemma 2b,'' Hugging Face, 2025, accessed: 2025-10-01. [Online]. Available: \url{https://huggingface.co/google/gemma-2-2b}
\BIBentrySTDinterwordspacing

\bibitem{smollm2}
\BIBentryALTinterwordspacing
H.~Face, ``Smollm2-1.7b-instruct,'' Hugging Face, 2025, accessed: 2025-10-01. [Online]. Available: \url{https://huggingface.co/HuggingFaceTB/SmolLM2-1.7B-Instruct}
\BIBentrySTDinterwordspacing

\bibitem{stablelm2}
\BIBentryALTinterwordspacing
S.~AI, ``Stablelm-2-1.6b,'' Hugging Face, 2025, accessed: 2025-10-01. [Online]. Available: \url{https://huggingface.co/stabilityai/stablelm-2-1\_6b}
\BIBentrySTDinterwordspacing

\bibitem{qwen2_5}
\BIBentryALTinterwordspacing
U.~AI, ``Qwen2.5-1.5b,'' Hugging Face, 2025, accessed: 2025-10-01. [Online]. Available: \url{https://huggingface.co/unsloth/Qwen2.5-1.5B}
\BIBentrySTDinterwordspacing

\bibitem{deepseek-r1-distill}
\BIBentryALTinterwordspacing
DeepSeek-AI, ``Deepseek-r1-distill-qwen-1.5b,'' Hugging Face, 2025, accessed: 2025-10-01. [Online]. Available: \url{https://huggingface.co/deepseek-\\ai/DeepSeek-R1-Distill-Qwen-1.5B}
\BIBentrySTDinterwordspacing

\bibitem{osc_batchprocessing}
\BIBentryALTinterwordspacing
(2025) Monitoring and managing your job. Ohio Supercomputer Center. Accessed: 2025-10-01. [Online]. Available: \url{https://www.osc.edu/\\supercomputing/batch-processing-at-osc/monitoring-and-managing-your-job}
\BIBentrySTDinterwordspacing

\bibitem{princeton_jobpriority}
\BIBentryALTinterwordspacing
(2025) Job priority. Princeton Research Computing. Accessed: 2025-10-01. [Online]. Available: \url{https://researchcomputing.princeton.edu/\\support/knowledge-base/job-priority}
\BIBentrySTDinterwordspacing

\bibitem{alcf_queuescheduling}
\BIBentryALTinterwordspacing
(2025) Queue scheduling. Argonne Leadership Computing Facility (ALCF). Accessed: 2025-10-01. [Online]. Available: \url{https://docs.alcf.anl.gov/policies/queue-scheduling/}
\BIBentrySTDinterwordspacing

\bibitem{Deloitte_2025}
\BIBentryALTinterwordspacing
{Deloitte Insights}, ``2025 global semiconductor industry outlook,'' February 2025. [Online]. Available: \url{https://www.deloitte.com/us/en/insights/industry/technology/technology-media-telecom-outlooks/semiconductor-industry-outlook.html}
\BIBentrySTDinterwordspacing

\bibitem{pilz2025trends}
K.~F. Pilz, J.~Sanders, R.~Rahman, and L.~Heim, ``Trends in ai supercomputers,'' \emph{arXiv preprint arXiv:2504.16026}, 2025.

\bibitem{kachris2025survey}
C.~Kachris, ``A survey on hardware accelerators for large language models,'' \emph{Applied Sciences}, vol.~15, no.~2, p. 586, 2025.

\bibitem{luo2024scheduling}
Y.~Luo, Q.~Wang, S.~Shi, J.~Lai, S.~Qi, J.~Zhang, and X.~Wang, ``Scheduling deep learning jobs in multi-tenant gpu clusters via wise resource sharing,'' in \emph{2024 IEEE/ACM 32nd International Symposium on Quality of Service (IWQoS)}.\hskip 1em plus 0.5em minus 0.4em\relax IEEE, 2024, pp. 1--10.

\bibitem{liang2024resource}
F.~Liang, Z.~Zhang, H.~Lu, C.~Li, V.~Leung, Y.~Guo, and X.~Hu, ``Resource allocation and workload scheduling for large-scale distributed deep learning: A survey,'' \emph{arXiv preprint arXiv:2406.08115}, 2024.

\bibitem{ding2023mirage}
Q.~Ding, P.~Zheng, S.~Kudari, S.~Venkataraman, and Z.~Zhang, ``Mirage: Towards low-interruption services on batch gpu clusters with reinforcement learning,'' in \emph{Proceedings of the International Conference for High Performance Computing, Networking, Storage and Analysis}, 2023, pp. 1--13.

\bibitem{jeon2018multi}
M.~Jeon, S.~Venkataraman, J.~Qian, A.~Phanishayee, W.~Xiao, and F.~Yang, ``Multi-tenant gpu clusters for deep learning workloads: Analysis and implications,'' \emph{Technical report, Microsoft Research}, 2018.

\bibitem{xiao2020antman}
W.~Xiao, S.~Ren, Y.~Li, Y.~Zhang, P.~Hou, Z.~Li, Y.~Feng, W.~Lin, and Y.~Jia, ``$\{$AntMan$\}$: Dynamic scaling on $\{$GPU$\}$ clusters for deep learning,'' in \emph{14th USENIX Symposium on Operating Systems Design and Implementation (OSDI 20)}, 2020, pp. 533--548.

\bibitem{jeon2019analysis}
M.~Jeon, S.~Venkataraman, A.~Phanishayee, J.~Qian, W.~Xiao, and F.~Yang, ``Analysis of $\{$Large-Scale$\}$$\{$Multi-Tenant$\}$$\{$GPU$\}$ clusters for $\{$DNN$\}$ training workloads,'' in \emph{2019 USENIX Annual Technical Conference (USENIX ATC 19)}, 2019, pp. 947--960.

\bibitem{nersc_policy}
\BIBentryALTinterwordspacing
(2025) Queues and charges. NERSC. Accessed: 2025-10-02. [Online]. Available: \url{https://docs.nersc.gov/jobs/policy/\#qos-cost-factor-charge-multipliers-and-discounts}
\BIBentrySTDinterwordspacing

\bibitem{ncar_jobs}
\BIBentryALTinterwordspacing
(2025) Job premption with pbs. NCAR HPC. Accessed: 2025-10-02. [Online]. Available: \url{https://ncar-hpc-docs.readthedocs.io/en/latest/pbs/preemption/\#charging-and-allocations}
\BIBentrySTDinterwordspacing

\bibitem{umd_queues}
\BIBentryALTinterwordspacing
(2025) Available hpc partitions. University of Maryland High-Performance Computing Center. Accessed: 2025-10-02. [Online]. Available: \url{https://hpcc.umd.edu/kb/queues/\#scavenger-partition}
\BIBentrySTDinterwordspacing

\bibitem{utah_preemption}
\BIBentryALTinterwordspacing
(2025) Atomatic restarting of preemptable jobs. Center for High Performance Computing at the University of Utah. Accessed: 2025-10-02. [Online]. Available: \url{https://www.chpc.utah.edu/documentation/software/slurm-job-preemption.php\#Automatic\%20Restarting\%20of\%20Preemptable\%20Jobs}
\BIBentrySTDinterwordspacing

\bibitem{fnal_slurm}
\BIBentryALTinterwordspacing
(2025) Slurm job scheduler. Fermilab. Accessed: 2025-10-02. [Online]. Available: \url{https://computing.fnal.gov/wilsoncluster/slurm-job-scheduler/}
\BIBentrySTDinterwordspacing

\bibitem{buffalo_jobs}
\BIBentryALTinterwordspacing
(2025) Slurm directives, partitions \& qos. Center for Computational Research at the University at Buffalo. Accessed: 2025-10-02. [Online]. Available: \url{https://docs.ccr.buffalo.edu/en/latest/hpc/jobs/\#slurm-directives-partitions-qos}
\BIBentrySTDinterwordspacing

\bibitem{qiu2023aware}
H.~Qiu, W.~Mao, C.~Wang, H.~Franke, A.~Youssef, Z.~T. Kalbarczyk, T.~Ba{\c{s}}ar, and R.~K. Iyer, ``$\{$AWARE$\}$: Automate workload autoscaling with reinforcement learning in production cloud systems,'' in \emph{2023 USENIX Annual Technical Conference (USENIX ATC 23)}, 2023, pp. 387--402.

\bibitem{zou2024optscaler}
D.~Zou, W.~Lu, Z.~Zhu, X.~Lu, J.~Zhou, X.~Wang, K.~Liu, K.~Wang, R.~Sun, and H.~Wang, ``Optscaler: A collaborative framework for robust autoscaling in the cloud,'' \emph{Proceedings of the VLDB Endowment}, vol.~17, no.~12, pp. 4090--4103, 2024.

\bibitem{augustyn2024tuning}
D.~R. Augustyn, {\L}.~Wyci{\'s}lik, and M.~Sojka, ``Tuning a kubernetes horizontal pod autoscaler for meeting performance and load demands in cloud deployments,'' \emph{Applied Sciences}, vol.~14, no.~2, p. 646, 2024.

\bibitem{catillo2023survey}
M.~Catillo, U.~Villano, and M.~Rak, ``A survey on auto-scaling: how to exploit cloud elasticity,'' \emph{International Journal of Grid and Utility Computing}, vol.~14, no.~1, pp. 37--50, 2023.

\bibitem{goulart2023checkpointing}
H.~Goulart, A.~Franco, and O.~Mendizabal, ``Checkpointing techniques in distributed systems: A synopsis of diverse strategies over the last decades,'' in \emph{Workshop de Testes e Toler{\^a}ncia a Falhas (WTF)}.\hskip 1em plus 0.5em minus 0.4em\relax SBC, 2023, pp. 15--28.

\bibitem{islam2012mcrengine}
T.~Z. Islam, K.~Mohror, S.~Bagchi, A.~Moody, B.~R. De~Supinski, and R.~Eigenmann, ``Mcrengine: A scalable checkpointing system using data-aware aggregation and compression,'' in \emph{SC'12: Proceedings of the International Conference on High Performance Computing, Networking, Storage and Analysis}.\hskip 1em plus 0.5em minus 0.4em\relax IEEE, 2012, pp. 1--11.

\bibitem{siachamis2024checkmate}
G.~Siachamis, K.~Psarakis, M.~Fragkoulis, A.~Van~Deursen, P.~Carbone, and A.~Katsifodimos, ``Checkmate: Evaluating checkpointing protocols for streaming dataflows,'' in \emph{2024 IEEE 40th international conference on data engineering (ICDE)}.\hskip 1em plus 0.5em minus 0.4em\relax IEEE, 2024, pp. 4030--4043.

\bibitem{babuji2019parsl}
Y.~Babuji, A.~Woodard, Z.~Li, D.~S. Katz, B.~Clifford, R.~Kumar, L.~Lacinski, R.~Chard, J.~M. Wozniak, I.~Foster \emph{et~al.}, ``Parsl: Pervasive parallel programming in python,'' in \emph{Proceedings of the 28th International Symposium on High-Performance Parallel and Distributed Computing}, 2019, pp. 25--36.

\bibitem{sly2023taskvine}
B.~Sly-Delgado, T.~S. Phung, C.~Thomas, D.~Simonetti, A.~Hennessee, B.~Tovar, and D.~Thain, ``Taskvine: Managing in-cluster storage for high-throughput data intensive workflows,'' in \emph{Proceedings of the SC'23 Workshops of the International Conference on High Performance Computing, Network, Storage, and Analysis}, 2023, pp. 1978--1988.

\bibitem{phung2023maximizing}
T.~S. Phung, B.~Clifford, K.~Chard, and D.~Thain, ``Maximizing data utility for hpc python workflow execution,'' in \emph{Proceedings of the SC'23 Workshops of the International Conference on High Performance Computing, Network, Storage, and Analysis}, 2023, pp. 637--640.

\bibitem{phung2024accelerating}
T.~S. Phung, C.~Thomas, L.~Ward, K.~Chard, and D.~Thain, ``Accelerating function-centric applications by discovering, distributing, and retaining reusable context in workflow systems,'' in \emph{Proceedings of the 33rd International Symposium on High-Performance Parallel and Distributed Computing}, 2024, pp. 122--134.

\bibitem{phung2024adaptive}
T.~S. Phung and D.~Thain, ``Adaptive task-oriented resource allocation for large dynamic workflows on opportunistic resources,'' in \emph{2024 IEEE International Parallel and Distributed Processing Symposium (IPDPS)}.\hskip 1em plus 0.5em minus 0.4em\relax IEEE, 2024, pp. 300--311.

\bibitem{zhang2023towards}
X.~Zhang and W.~Gao, ``Towards llm-based fact verification on news claims with a hierarchical step-by-step prompting method,'' \emph{arXiv preprint arXiv:2310.00305}, 2023.

\bibitem{gunjal2024molecular}
A.~Gunjal and G.~Durrett, ``Molecular facts: Desiderata for decontextualization in llm fact verification,'' \emph{arXiv preprint arXiv:2406.20079}, 2024.

\bibitem{Thorne18Fever}
J.~Thorne, A.~Vlachos, C.~Christodoulopoulos, and A.~Mittal, ``{FEVER}: a large-scale dataset for fact extraction and {VERification},'' in \emph{NAACL-HLT}, 2018.

\bibitem{allal2025smollm2}
L.~B. Allal, A.~Lozhkov, E.~Bakouch, G.~M. Bl{\'a}zquez, G.~Penedo, L.~Tunstall, A.~Marafioti, H.~Kydl{\'\i}{\v{c}}ek, A.~P. Lajar{\'\i}n, V.~Srivastav \emph{et~al.}, ``Smollm2: When smol goes big--data-centric training of a small language model,'' \emph{arXiv preprint arXiv:2502.02737}, 2025.

\bibitem{thain2005distributed}
D.~Thain, T.~Tannenbaum, and M.~Livny, ``Distributed computing in practice: the condor experience,'' \emph{Concurrency and computation: practice and experience}, vol.~17, no. 2-4, pp. 323--356, 2005.

\bibitem{shaffer2017taming}
T.~Shaffer and D.~Thain, ``Taming metadata storms in parallel filesystems with metafs,'' in \emph{Proceedings of the 2nd Joint International Workshop on Parallel Data Storage \& Data Intensive Scalable Computing Systems}, 2017, pp. 25--30.

\bibitem{panasas}
B.~Welch, M.~Unangst, Z.~Abbasi, G.~A. Gibson, B.~Mueller, J.~Small, J.~Zelenka, and B.~Zhou, ``Scalable performance of the panasas parallel file system.'' in \emph{FAST}, vol.~8, 2008, pp. 1--17.

\bibitem{conda}
A.~Inc., ``Anaconda software distribution,'' https://docs.anaconda.com/, 2020.

\bibitem{GoogleCloudSpotVMs}
\BIBentryALTinterwordspacing
(2025) Spot vms. Google Cloud. [Online]. Available: \url{https://cloud.google.com/solutions/spot-vms}
\BIBentrySTDinterwordspacing

\bibitem{AWSEC2Spot}
\BIBentryALTinterwordspacing
(2025) Amazon ec2 spot instances. {Amazon Web Services (AWS)}. [Online]. Available: \url{https://aws.amazon.com/ec2/spot/}
\BIBentrySTDinterwordspacing

\bibitem{AzureSpotVMs}
\BIBentryALTinterwordspacing
(2025) Spot virtual machines. {Microsoft Azure}. [Online]. Available: \url{https://azure.microsoft.com/en-us/products/virtual-machines/spot}
\BIBentrySTDinterwordspacing

\bibitem{mao2025skyserve}
Z.~Mao, T.~Xia, Z.~Wu, W.-L. Chiang, T.~Griggs, R.~Bhardwaj, Z.~Yang, S.~Shenker, and I.~Stoica, ``Skyserve: Serving ai models across regions and clouds with spot instances,'' in \emph{Proceedings of the Twentieth European Conference on Computer Systems}, 2025, pp. 159--175.

\bibitem{miao2024spotserve}
X.~Miao, C.~Shi, J.~Duan, X.~Xi, D.~Lin, B.~Cui, and Z.~Jia, ``Spotserve: Serving generative large language models on preemptible instances,'' in \emph{Proceedings of the 29th ACM International Conference on Architectural Support for Programming Languages and Operating Systems, Volume 2}, 2024, pp. 1112--1127.

\bibitem{AWSSpotTermination}
\BIBentryALTinterwordspacing
(2025) Spot instance interruption notices. {Amazon Web Services (AWS)}. [Online]. Available: \url{https://docs.aws.amazon.com/AWSEC2/latest/UserGuide/spot-instance-termination-notices.html}
\BIBentrySTDinterwordspacing

\bibitem{GoogleCloudSpot}
\BIBentryALTinterwordspacing
(2025) Spot vms. Google Cloud. [Online]. Available: \url{https://cloud.google.com/compute/docs/instances/spot}
\BIBentrySTDinterwordspacing

\bibitem{AzureSpotVMsDoc}
\BIBentryALTinterwordspacing
(2025) Spot virtual machines. {Microsoft Azure}. [Online]. Available: \url{https://learn.microsoft.com/en-us/azure/virtual-machines/spot-vms}
\BIBentrySTDinterwordspacing

\bibitem{leviathan2023fast}
Y.~Leviathan, M.~Kalman, and Y.~Matias, ``Fast inference from transformers via speculative decoding,'' in \emph{International Conference on Machine Learning}.\hskip 1em plus 0.5em minus 0.4em\relax PMLR, 2023, pp. 19\,274--19\,286.

\bibitem{spector2023accelerating}
B.~Spector and C.~Re, ``Accelerating llm inference with staged speculative decoding,'' \emph{arXiv preprint arXiv:2308.04623}, 2023.

\bibitem{chen2024cascade}
Z.~Chen, X.~Yang, J.~Lin, C.~Sun, K.~Chang, and J.~Huang, ``Cascade speculative drafting for even faster llm inference,'' \emph{Advances in Neural Information Processing Systems}, vol.~37, pp. 86\,226--86\,242, 2024.

\bibitem{svirschevski2024specexec}
R.~Svirschevski, A.~May, Z.~Chen, B.~Chen, Z.~Jia, and M.~Ryabinin, ``Specexec: Massively parallel speculative decoding for interactive llm inference on consumer devices,'' \emph{Advances in Neural Information Processing Systems}, vol.~37, pp. 16\,342--16\,368, 2024.

\bibitem{kwon2023efficient}
W.~Kwon, Z.~Li, S.~Zhuang, Y.~Sheng, L.~Zheng, C.~H. Yu, J.~Gonzalez, H.~Zhang, and I.~Stoica, ``Efficient memory management for large language model serving with pagedattention,'' in \emph{Proceedings of the 29th Symposium on Operating Systems Principles}, 2023, pp. 611--626.

\bibitem{lin2024infinite}
B.~Lin, C.~Zhang, T.~Peng, H.~Zhao, W.~Xiao, M.~Sun, A.~Liu, Z.~Zhang, L.~Li, X.~Qiu \emph{et~al.}, ``Infinite-llm: Efficient llm service for long context with distattention and distributed kvcache,'' \emph{arXiv preprint arXiv:2401.02669}, 2024.

\bibitem{fu2024serverlessllm}
Y.~Fu, L.~Xue, Y.~Huang, A.-O. Brabete, D.~Ustiugov, Y.~Patel, and L.~Mai, ``$\{$ServerlessLLM$\}$:$\{$Low-Latency$\}$ serverless inference for large language models,'' in \emph{18th USENIX Symposium on Operating Systems Design and Implementation (OSDI 24)}, 2024, pp. 135--153.

\bibitem{turilli2019middleware}
M.~Turilli, V.~Balasubramanian, A.~Merzky, I.~Paraskevakos, and S.~Jha, ``Middleware building blocks for workflow systems,'' \emph{Computing in Science \& Engineering}, vol.~21, no.~4, pp. 62--75, 2019.

\bibitem{zheng2017deploying}
C.~Zheng, B.~Tovar, and D.~Thain, ``Deploying high throughput scientific workflows on container schedulers with makeflow and mesos,'' in \emph{2017 17th IEEE/ACM International Symposium on Cluster, Cloud and Grid Computing (CCGRID)}.\hskip 1em plus 0.5em minus 0.4em\relax IEEE, 2017, pp. 130--139.

\bibitem{phung2021not}
T.~S. Phung, L.~Ward, K.~Chard, and D.~Thain, ``Not all tasks are created equal: Adaptive resource allocation for heterogeneous tasks in dynamic workflows,'' in \emph{2021 IEEE Workshop on Workflows in Support of Large-Scale Science (WORKS)}.\hskip 1em plus 0.5em minus 0.4em\relax IEEE, 2021, pp. 17--24.

\bibitem{rocklin2015dask}
M.~Rocklin \emph{et~al.}, ``Dask: Parallel computation with blocked algorithms and task scheduling.'' in \emph{SciPy}, 2015, pp. 126--132.

\bibitem{moritz2018ray}
P.~Moritz, R.~Nishihara, S.~Wang, A.~Tumanov, R.~Liaw, E.~Liang, M.~Elibol, Z.~Yang, W.~Paul, M.~I. Jordan \emph{et~al.}, ``Ray: A distributed framework for emerging $\{$AI$\}$ applications,'' in \emph{13th USENIX symposium on operating systems design and implementation (OSDI 18)}, 2018, pp. 561--577.

\end{thebibliography}

\end{document}